\documentclass[a4paper,12pt]{article}
\usepackage{amsmath}
\usepackage{amssymb}
\usepackage{amsfonts}
\usepackage{bbm} 
\usepackage{fleqn} 

\usepackage[footnotesize]{caption}
\usepackage{graphicx} 
\usepackage{braket} 
\usepackage{mathrsfs}
\usepackage[center,footnotesize,hang]{subfigure}
\usepackage[bbgreekl]{MyBbol}
\usepackage{cite}
\usepackage{pstricks}

\newcommand{\CenterEps}[2][1]{\ensuremath{\vcenter{\hbox{\includegraphics[scale=#1]{#2.eps}}}}}

\def\D{\mathrm{d}} 
\def\I{\mathrm{i}}
\def\ChargeC{\mathrm{C}}                
\newcommand{\SuperField}[1]{\bbsymbol{#1}}  
\DeclareMathOperator{\re}{Re}
\DeclareMathOperator{\im}{Im}

\DeclareMathOperator{\diag}{diag}

\allowdisplaybreaks[1]

\begin{document}

\begin{titlepage}

\ \vspace*{-15mm}
\begin{flushright}
TUM-HEP-510/03\\
DESY 03-065
\end{flushright}
\vspace*{5mm}

\begin{center}
{\Large\sffamily\bfseries 
Running Neutrino Masses, Mixings and CP Phases: Analytical Results
and Phenomenological Consequences}
\\[12mm]
Stefan Antusch\footnote{E-mail: \texttt{santusch@ph.tum.de}},
J\"{o}rn Kersten\footnote{E-mail: \texttt{jkersten@ph.tum.de}},
Manfred Lindner\footnote{E-mail: \texttt{lindner@ph.tum.de}}
\\
{\small\it
Physik-Department T30, 
Technische Universit\"{a}t M\"{u}nchen\\ 
James-Franck-Stra{\ss}e,
85748 Garching, Germany
}
\\[3mm]
Michael Ratz\footnote{E-mail: \texttt{mratz@mail.desy.de}}\\
{\small\it
Deutsches Elektronensynchrotron DESY\\ 
22603 Hamburg, Germany
}
\end{center}
\vspace*{2.0cm}

\begin{abstract}
\noindent
We derive simple analytical formulae for the renormalization group 
running of neutrino masses, leptonic mixing angles and CP phases,
which allow an easy understanding of the running. Particularly for 
a small angle $\theta_{13}$ the expressions become very compact,
even when non-vanishing CP phases are present. 
Using these equations we investigate:
(i) the influence of Dirac and Majorana phases on the evolution of all parameters,
(ii) the implications of running neutrino parameters for leptogenesis,
(iii) changes of the mass bounds from WMAP and neutrinoless double 
$\beta$ decay experiments, relevant for high-energy mass models,
(iv) the size of radiative corrections to $\theta_{13}$ and $\theta_{23}$
and implications for future precision measurements.

\end{abstract}

\end{titlepage}

\newpage
\setcounter{footnote}{0}

\section{Introduction}

The Standard Model (SM) agrees very well with experiments and the 
only solid evidence for new physics consists in the observation 
of neutrino masses. Compared to quarks and charged leptons they
are tiny, for which the see-saw mechanism
\cite{Yanagida:1980,Glashow:1979vf,Gell-Mann:1980vs,Mohapatra:1980ia} 
provides an attractive explanation. The parameters which enter into 
the neutrino mass matrix usually stem from model predictions at high 
energy scales, such as the scale \(M_\mathrm{GUT}\) of grand unification. 
The measurements and bounds for neutrino masses and lepton 
mixings, on the other hand, determine the parameters at low energy. 
The high- and low-energy parameters are related by the renormalization 
group (RG) evolution, so that low-energy data yield only indirect 
restrictions for mass models or other high-energy mechanisms like 
leptogenesis \cite{Fukugita:1986hr}. 
It is well-known that the model independent RG evolution between low energy 
and the lowest see-saw scale can have large effects on the leptonic mixing 
angles and on the mass squared differences,
in particular if the neutrinos have quasi-degenerate masses 
\cite{Tanimoto:1995bf,Ellis:1999my,Casas:1999tp,Casas:1999ac,%
Chankowski:1999xc,Casas:1999tg,Balaji:2000au,Haba:2000tx,Miura:2000bj,%
Chankowski:2000fp,Chen:2001gk,Chankowski:2001mx,Parida:2002gz,%
Dutta:2002nq,Miura:2002nz,Bhattacharyya:2002aq,Joshipura:2002xa,%
Frigerio:2002in}.
RG effects may even serve as an explanation for the discrepancy between 
the mixings in the quark and the lepton sector \cite{Mohapatra:2003tw}.

The RG equations (RGEs) for the neutrino mass operator
and for all the other parameters of the theory have to be solved 
simultaneously. The mixing angles, phases and mass eigenvalues can then 
be extracted from the evolved mass matrices. Both steps are, however, 
non-trivial and can only be performed numerically in practice.
In order to determine the change of the parameters under the RG flow
in a qualitative and, to a reasonable accuracy, also quantitative
way, it is useful to derive analytical formulae for the running of the
masses, mixing angles and phases. This was done in \cite{Chankowski:1999xc}
assuming CP conservation and in \cite{Casas:1999tg} for the general
case. 
We modify the derivation of \cite{Casas:1999tg} by a step which simplifies the
formulae that arise after explicitly writing out the dependence on the
mixing parameters.  These results are exact, and they make it
easier to derive simple approximations in the limit of small $\theta_{13}$.
These approximations are very useful in understanding the RG evolution of 
the phases and the phase dependence of the evolution of other parameters.
For example, we find that the phases show significant running. Consequently, 
vanishing phases at low energy appear unnatural unless exact CP conservation 
is a boundary condition at high energy, which seems unlikely, since the 
CP phase in the quark sector is sizable. The presence of CP phases at low energies 
has significant impact on observations
\cite{Dick:1999ed,Frigerio:2002rd,Frigerio:2002fb}.

The outline for the paper is: 
In Sec.~\ref{sec:RGEvolution} we present analytical formulae 
for the RG evolution of the neutrino masses, leptonic mixing 
angles and phases, where an expansion in the small angle 
\(\theta_{13}\) is performed. This leads to very simple and in most 
cases accurate formulae which are compared with numerical results. 
Sec.~\ref{sec:Applications} is devoted to phenomenological 
consequences for leptogenesis, the WMAP bound, the 
effective neutrino mass relevant for neutrinoless double
beta decay and precision measurements of $\theta_{13}$
and $\theta_{23}$.

\section{RG Evolution of Leptonic Mixing Parameters and Neutrino Masses}
\label{sec:RGEvolution}

\begin{figure}[!h]
\begin{center}
  $
  \CenterEps{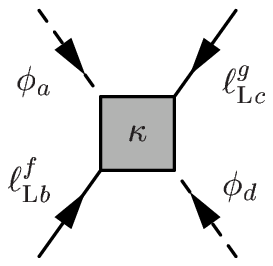}
  \xrightarrow[\mbox{{\small breaking}}]{\mbox{{\small EW symmetry}}}
  \CenterEps{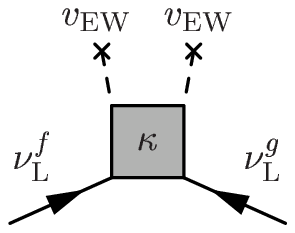}
  $
 \caption{ \label{fig:KappaVertex} Vertex from the dimension 5 operator 
 which yields a Majorana mass matrix for the light neutrinos.}
\end{center}
\end{figure}

In this study, we will focus on neutrino masses which can be described
by the lowest-dimensional neutrino mass operator compatible with 
the gauge symmetries of the SM. This operator reads in the SM
\begin{equation}\label{eq:Kappa:Babu:1993:1}
 \mathscr{L}_{\kappa} 
 =\frac{1}{4} 
 \kappa_{gf} \, \overline{\ell_\mathrm{L}^\mathrm{C}}^g_c\varepsilon^{cd} \phi_d\, 
 \, \ell_{\mathrm{L}b}^{f}\varepsilon^{ba}\phi_a  
  +\text{h.c.} 
  \;,
\end{equation}
and in its minimal supersymmetric extension, the MSSM,
\begin{equation}\label{eq:Kappa-MSSM-s}
\mathscr{L}_{\kappa}^{\mathrm{MSSM}} 
\,=\, \mathscr{W}_\mathrm{\kappa} \big|_{\theta\theta}   +\text{h.c.}
= -\tfrac{1}{4} 
  {\kappa}^{}_{gf} \, \SuperField{l}^{g}_c\varepsilon^{cd}
 \SuperField{h}^{(2)}_d\, 
 \, \SuperField{l}_{b}^{f}\varepsilon^{ba} \SuperField{h}^{(2)}_a 
 \big|_{\theta\theta}   +\text{h.c.} \;.
\end{equation}
$\kappa_{gf}$ has mass dimension $-1$ and is symmetric
under interchange of the generation indices $f$ and $g$, 
$\varepsilon$ is the totally antisymmetric tensor in 
2 dimensions, and $\ell_\mathrm{L}^\ChargeC$ is the charge 
conjugate of a lepton doublet. 
\(a,b,c,d \in \{1,2\}\) are $\mathrm{SU}(2)_\mathrm{L}$ indices.  
The double-stroke letters \(\SuperField{l}\) and \(\SuperField{h}\)
denote lepton doublets and the up-type Higgs superfield in the MSSM.
After electroweak (EW) symmetry breaking, a Majorana neutrino mass 
matrix proportional to $\kappa$ emerges as illustrated in Fig.~\ref{fig:KappaVertex}.  

The above mass operator provides a rather model-independent way to
introduce neutrino masses as there are many possibilities to realize it 
radiatively or at tree-level within a renormalizable theory 
(see e.g.~\cite{Ma:1998dn}). The tree-level
realizations from integrating out heavy singlet fermions and/or Higgs 
triplets naturally appear for instance in left-right-symmetric
extensions of the SM or MSSM and
are usually referred to as type I and type II see-saw mechanisms.

The energy dependence of the effective neutrino mass matrix below the
scale where the operator is generated (which we will call \(M_1\) 
in the following) is described by its RGE.  At the one-loop level, 
this equation is given by
\cite{Chankowski:1993tx,Babu:1993qv,Antusch:2001ck,Antusch:2001vn}
\begin{equation}\label{eq:BetaKappa}
 16\pi^2 \, \frac{\D\kappa}{\D t} 
 \,=\,
 C\,(Y_e^\dagger Y_e)^T\,\kappa+C\,\kappa\,(Y_e^\dagger Y_e) + \alpha\,\kappa\;,
\end{equation}
where $t=\ln(\mu/\mu_0)$ and $\mu$ is the renormalization scale\footnote{In 
the MSSM, the RGE is known at two-loop \cite{Antusch:2002ek}. In this study,
we will, however, focus on the one-loop equation.}
and where
\begin{eqnarray}
        C &=& 1 \hphantom{-\frac{3}{2}} \;\; \text{in the MSSM}\;,
\nonumber\\
        C &=& -\frac{3}{2} \hphantom{1} \;\; \text{in the SM}\;.
\end{eqnarray}
In the SM and in the MSSM, \(\alpha\) reads
\begin{subequations}
\begin{eqnarray}
        \alpha_\mathrm{SM}
& = &
        -3 g_2^2 + 2 (y_\tau^2+y_\mu^2+y_e^2) +
        6 \left( y_t^2 + y_b^2 + y_c^2 + y_s^2 + y_d^2 + y_u^2 \right) 
        + \lambda \;,
\\
        \alpha_\mathrm{MSSM}
& = &
 -\frac{6}{5} g_1^2 - 6 g_2^2 + 6 \left( y_t^2 + y_c^2 + y_u^2 \right)
 \;.
\end{eqnarray}
\end{subequations}
Here \(Y_f\) (\(f\in\{e,d,u\}\)) represent the Yukawa coupling matrices of 
the charged leptons, down- and up-type quarks, respectively, \(g_i\) denote
the gauge couplings\footnote{We are using GUT charge normalization for $g_1$.} 
and \(\lambda\) the Higgs self-coupling in the SM.  We work in the basis
where $Y_e$ is diagonal.

The parameters of interest are the masses, which are proportional to the
eigenvalues of \(\kappa\) and defined to be non-negative, as well as the
mixing angles and physical phases of the MNS matrix \cite{Maki:1962mu} 
\begin{eqnarray}\label{eq:StandardParametrizationUMNS}
 U_\mathrm{MNS} & = & V(\theta_{12},\theta_{13},\theta_{23},\delta) \, 
 \diag(e^{-\I\varphi_1/2},e^{-\I\varphi_2/2},1)
 \;,
\end{eqnarray}
which diagonalizes \(\kappa\) in this basis. $V$ is the leptonic
analogon to the CKM matrix in the quark sector.  The
parametrization we use will be explained in more detail in 
App.~\ref{app:MixingParameters}.
Currently, we learn from experiments that there occur two oscillations
with mass squared differences \(\Delta m^2_\mathrm{sol}\) and
\(\Delta m^2_\mathrm{atm}\) and corresponding mixing angles
\(\theta_{12}\) and \(\theta_{23}\), respectively. For the third
mixing angle \(\theta_{13}\) and the absolute scale of light
neutrino masses, there are only upper bounds at the moment
(see Tab.~\ref{tab:ExpData} for the present status).
\begin{table}[hbt]
\begin{center}
\begin{tabular}{l|ccc}
&$\vphantom{\sqrt{\big|}}$ Best-fit value & Range (for \(\theta_{ij} \in [0^\circ,45^\circ]\)) & C.L. \\
  \hline
$\vphantom{\sqrt{\big|}}$\(\theta_{12}\) [\({\:}^\circ\)]& \(32.6\) & \(25.6 - 42.0\) & 
\(99 \% 
\;(3\sigma)\)\\  
$\vphantom{\sqrt{\big|}}$\(\theta_{23}\) [\({\:}^\circ\)]& \(45.0\) & \(33.2 - 45.0\) &    
\(99 \%
\;(3 \sigma)\)\\
$\vphantom{\sqrt{\big|}}$\(\theta_{13}\) [\({\:}^\circ\)]& \(-\) &    
\(0.0-  9.2\) &     \(90 \% \)\\
$\vphantom{\sqrt{\big|}}$\(\Delta m^2_{\mathrm{sol}}\) [eV\(^2\)] & \(7.3 \cdot 10^{-5}\)&
\(4\cdot 10^{-5} - 2.8\cdot 10^{-4}\) & \(99 \%
\;(3 \sigma)\) \\
$\vphantom{\sqrt{\big|}}$\(|\Delta m^2_{\mathrm{atm}}|\) [eV\(^2\)] & \(2.5 \cdot 10^{-3}\)&
\(1.2\cdot 10^{-3} - 5\cdot 10^{-3}\) & \(99 \%
\;(3 \sigma)\)\hfill
\end{tabular}
\caption{
 Experimental data for the neutrino mixing angles and mass squared
 differences. 
 For the solar angle $\theta_{12}$ and the solar mass squared
 difference, the LMA solution as confirmed by KamLAND is shown.
 The results stem from the analysis \cite{deHolanda:2002iv} of the recent KamLAND 
 and the SNO data, the Super-Kamiokande atmospheric data
 \cite{Toshito:2001dk} and the CHOOZ experiment \cite{Apollonio:1999ae}. 
}\label{tab:ExpData}
\vspace{-0.35cm}
\end{center}
\end{table}

\subsection{The Analytical Formulae}\label{sec:AnalyticFormulae}

In this section, we present explicit RGEs for the physical parameters.
They determine the slope of the RG evolution at a
given energy scale and thus yield an insight into the RG behavior.
The derivation will be discussed in App.~\ref{sec:DerivRGEs}.
Note that a naive 
linear interpolation, i.e.\  assuming the right-hand sides of the
equations to be constant, will not always give the correct RG evolution.
As we will show later, this is mainly due to large changes of
$\theta_{12}$ and the mass squared differences.
In the following, we will neglect ${y_e}$ and ${y_\mu}$ against \(y_\tau\) and
introduce the abbreviation
\begin{equation}
        \zeta
        \,:=\,
        \frac{\Delta m^2_\mathrm{sol}}{\Delta m^2_\mathrm{atm}} \;,
\end{equation}
whose LMA best-fit value is about 0.03. In order to keep the expressions 
short, we will only show the leading terms in an expansion in the small
angle \(\theta_{13}\) for the mixing parameters.
In almost all cases they are sufficient for understanding the features
of the RG evolution.%
\footnote{The exact formulae, from which we have derived the analytical 
approximations presented here, can be obtained from the web page
 \texttt{http://www.ph.tum.de/\textasciitilde{}mratz/AnalyticFormulae/}.
}
In all cases except for the running of the Dirac phase \(\delta\), the limit
\(\theta_{13}\to0\) causes no difficulties, the subtleties
arising for \(\delta\) will be discussed in
Sec.~\ref{sec:RunningOfDelta}.
We furthermore define \(m_i(t):=v^2\,\kappa_i(t)/4\) with 
\(v=246\,\mathrm{GeV}\) in the SM or \(v=246\,\mathrm{GeV}\cdot \sin\beta\)
in the MSSM and, as usual, $\Delta m^2_\mathrm{sol} := m_2^2-m_1^2$ and 
$\Delta m^2_\mathrm{atm} := m_3^2-m_2^2$. Note that our formulae cannot be
applied if one of the mass squared differences vanishes.
For a discussion of RG effects in this case, see e.g.\ 
\cite{Ellis:1999my,Casas:1999tp,Casas:1999ac,Joshipura:2002xa,Joshipura:2002gr}.
With these conventions, we obtain the following analytical expressions
for the mixing angles:
\begin{eqnarray} \label{eq:AnalyticApproxT12}
\Dot{\theta}_{12}
& = &
        -\frac{C y_\tau^2}{32\pi^2} \,
        \sin 2\theta_{12} \, s_{23}^2\, 
        \frac{
      | {m_1}\, e^{\I \varphi_1} + {m_2}\, e^{\I  \varphi_2}|^2
     }{\Delta m^2_\mathrm{sol} }     
        + \mathscr{O}(\theta_{13}) \;,
         \label{eq:Theta12Dot}
\end{eqnarray} 
\begin{eqnarray} \label{eq:AnalyticApproxT13} 
\Dot{\theta}_{13}
& = & 
        \frac{C y_\tau^2}{32\pi^2} \, 
        \sin 2\theta_{12} \, \sin 2\theta_{23} \,
        \frac{m_3}{\Delta m^2_\mathrm{atm} \left( 1+\zeta \right)}
        \times
\nonumber\\
&& \quad \times
        \left[
         m_1 \cos(\varphi_1-\delta) -
         \left( 1+\zeta \right) m_2 \, \cos(\varphi_2-\delta) -
         \zeta m_3 \, \cos\delta
        \right]
        +       \mathscr{O}(\theta_{13}) \;,
 \label{eq:Theta13Dot}
\end{eqnarray}
\begin{eqnarray} \label{eq:AnalyticApproxT23} 
        \Dot{\theta}_{23}
& = &
        -\frac{C y_\tau^2}{32\pi^2} \, \sin 2\theta_{23} \,
        \frac{1}{\Delta m^2_\mathrm{atm}} 
        \left[
         c_{12}^2 \, |m_2\, e^{\I \varphi_2} + m_3|^2 +
         s_{12}^2 \, \frac{|m_1\, e^{\I \varphi_1} + m_3|^2}{1+\zeta}
        \right] 
        \nonumber\\
        & & {}
        + \mathscr{O}(\theta_{13}) \;.
 \label{eq:Theta23Dot}
\end{eqnarray}
Note that in order to apply Eq.~\eqref{eq:AnalyticApproxT13} to the case
$\theta_{13}=0$, where $\delta$ is undefined, the analytic continuation
of the latter, which will be given in Eq.~\eqref{eq:DeltaZeroT13}, has
to be inserted. 
The $\mathscr{O}(\theta_{13})$ terms in the above RGEs can become
important if $\theta_{13}$ is not too small and in particular if cancellations appear
in the leading terms.  For example, this is the case
for $|\varphi_1-\varphi_2|=\pi$ in \eqref{eq:Theta12Dot}, as we will
discuss below in more detail.
The RGE for the Dirac phase is given by
\begin{equation} \label{eq:DeltaPrimeWithNonZeroDelta}
 \Dot{\delta}
 \,=\,
 \frac{C y_\tau^2}{32\pi^2}
 \frac{\delta^{(-1)}}{\theta_{13}}
 +\frac{C y_\tau^2}{8\pi^2}\delta^{(0)}+\mathscr{O}(\theta_{13})\;,
\end{equation}
where
\begin{subequations}
\begin{eqnarray}
 \delta^{(-1)}
 & = &
 \sin2\theta_{12}\,\sin2\theta_{23}\,
 \frac{m_3}
        {\Delta m^2_\mathrm{atm} \left(1+\zeta\right)} \times
 \nonumber\\
 & &    
 \quad\times \left[
        m_1 \sin(\varphi_1-\delta) -
        \left( 1+\zeta \right) m_2\,\sin(\varphi_2-\delta) +
        \zeta m_3\,\sin\delta
 \right]
 \;,\\
  \delta^{(0)}
  & = &
  \frac{m_1 m_2\,s_{23}^2\,\sin(\varphi_1-\varphi_2)}{\Delta m^2_\mathrm{sol}}
  \nonumber\\
  & & {}
  +m_3\,s_{12}^2 \left[
        \frac{m_1\,\cos2\theta_{23}\,\sin\varphi_1}{\Delta m^2_\mathrm{atm}(1+\zeta)}
        +\frac{m_2\,c_{23}^2\,\sin(2\delta-\varphi_2)}{\Delta m^2_\mathrm{atm}}
  \right]
  \nonumber\\
  & & {}
  +m_3\,c_{12}^2 \left[
        \frac{m_1\,c_{23}^2\,\sin(2\delta-\varphi_1)}{\Delta m^2_\mathrm{atm}(1+\zeta)}
        +\frac{m_2\,\cos2\theta_{23}\,\sin\varphi_2}{\Delta m^2_\mathrm{atm}}
  \right]
  \;.
\end{eqnarray}
\end{subequations}
For the physical Majorana phases, we obtain
\begin{eqnarray}\label{eq:AnalyticApproxP1} 
        \Dot\varphi_1
& = & 
        \frac{C y_\tau^2}{4\pi^2} 
        \left\{ 
        m_3 \, \cos 2\theta_{23} \,
        \frac{
         m_1 s_{12}^2\,
         \sin\varphi_1 +
         \left( 1+\zeta \right) m_2 \, c_{12}^2 
         \, \sin\varphi_2 }
         { \Delta m^2_\mathrm{atm} \left( 1+\zeta \right) }
        \right. 
\nonumber\\
&& \hphantom{   \frac{C y_\tau^2}{4\pi^2} \left\{ \right.}      
        \left. {}
        +
        \frac{ m_1 m_2 \, c_{12}^2\,s_{23}^2\,
         \sin(\varphi_1-\varphi_2) }
         { \Delta m^2_\mathrm{sol} }
        \right\} +
        \mathscr{O}(\theta_{13}) \;,
        \label{eq:Phi1Dot}
\\
        \Dot\varphi_2                   \label{eq:AnalyticApproxP2} 
& = &
        \frac{C y_\tau^2}{4\pi^2}
        \left\{ 
        m_3 \, \cos 2\theta_{23} \,
        \frac{
         m_1 s_{12}^2\,
         \sin\varphi_1 +
         \left( 1+\zeta \right) m_2 \, c_{12}^2
         \sin\varphi_2 }
         { \Delta m^2_\mathrm{atm} \left( 1+\zeta \right) }
        \right. 
\nonumber\\
&& \hphantom{   \frac{C y_\tau^2}{4\pi^2} \left\{ \right.}      
        \left. {}
        +
        \frac{ m_1 m_2 \,
         s_{12}^2\,s_{23}^2\,
          \sin(\varphi_1-\varphi_2) }
         { \Delta m^2_\mathrm{sol} }
        \right\} +
        \mathscr{O}(\theta_{13}) \;.
        \label{eq:Phi2Dot}
\end{eqnarray}
We would like to emphasize that the above expressions do not contain
expansions in $\zeta$, i.e.\ their $\zeta$ dependence is exact.
In many cases, they  can be further simplified by
neglecting \(\zeta\) against 1 without losing much accuracy.
Note that singularities can appear in the $\mathscr{O}(\theta_{13})$-terms 
at points in parameter space where the phases are not well-defined. 
For the masses, the results for $y_e=y_\mu=0$ but arbitrary
$\theta_{13}$ are 
\begin{subequations}\label{eq:EvolutionOfMassEigenvalues}
\begin{eqnarray}
 16\pi^2\,\Dot{m}_{1}
 & = &
        \left[
        \alpha + C y_\tau^2 \left( 2 s_{12}^2 \, s_{23}^2 + F_1 \right)
        \right] m_1 \;,
\\
 16\pi^2\,\Dot{m}_2
 & = &
        \left[
        \alpha + C y_\tau^2 \left( 2 c_{12}^2 \, s_{23}^2 + F_2 \right)
        \right] m_2
 \;,
 \\
 16\pi^2\,\Dot{m}_3
 & = &
 \left[ \alpha
 +2 C y_\tau^2 \, c_{13}^2 \, c_{23}^2
 \right] m_3
 \;,
\end{eqnarray}
\end{subequations}
where \(F_1\) and \(F_2\) contain terms proportional to $\sin\theta_{13}$,
\begin{subequations}\label{eq:Fi}
\begin{eqnarray}
        F_1 &=&
        -s_{13} \, \sin 2\theta_{12} \, \sin 2\theta_{23} \, \cos\delta +
        2 s_{13}^2 \, c_{12}^2 \, c_{23}^2 \;,
\\
        F_2 &=&
    s_{13} \, \sin 2\theta_{12} \, \sin 2\theta_{23} \, \cos\delta +
        2 s_{13}^2 \, s_{12}^2 \, c_{23}^2 \;.
\end{eqnarray}
\end{subequations}
These formulae can be translated into RGEs for the mass squared
differences,
\begin{subequations}\label{eq:RunningDeltaM2s}
\begin{eqnarray}
        \! 8\pi^2 \, \frac{\D}{\D t} \Delta m_\mathrm{sol}^2 \,
& = & \!
        \alpha \, \Delta m_\mathrm{sol}^2 + 
        C y_\tau^2 \left[
         2 s_{23}^2 \left( m_2^2\,c_{12}^2 - m_1^2\,s_{12}^2 \right) +
         F_\mathrm{sol}
        \right]\; ,
        \label{eq:Dm2solDot}
\\
        8\pi^2 \, \frac{\D}{\D t} \Delta m_\mathrm{atm}^2
& = & \!
        \alpha \, \Delta m_\mathrm{atm}^2 +
        C y_\tau^2 \left[ 
         2 m_3^2 \, c_{13}^2 \, c_{23}^2 - 2 m_2^2 \, c_{12}^2 \, s_{23}^2 +
         F_\mathrm{atm} 
        \right]\; ,
        \label{eq:Dm2atmDot}
\end{eqnarray}
\end{subequations}
where 
\begin{subequations}\label{eq:FsolAndFatm}
\begin{eqnarray}
        F_\mathrm{sol}
& = &
        \left( m_1^2+m_2^2 \right)
        s_{13} \, \sin 2\theta_{12} \, \sin 2\theta_{23} \, \cos\delta
\nonumber\\
&& {} +
        2 s_{13}^2\, c_{23}^2 \left( m_2^2\,s_{12}^2-m_1^2\,c_{12}^2 \right)
        \;,\label{eq:Fsol}
\\
        F_\mathrm{atm} &=&
        -m_2^2 \,s_{13} \,\sin 2\theta_{12} \,\sin 2\theta_{23} \,\cos\delta
        -2 m_2^2 \, s_{13}^2 \, s_{12}^2 \, c_{23}^2 \;.\label{eq:Fatm}
\end{eqnarray}
\end{subequations}

\subsection{Generic Enhancement and Suppression Factors}

From Eqs.~\eqref{eq:Theta12Dot}--\eqref{eq:Phi2Dot} 
it follows that there are generic enhancement and suppression factors for the 
RG evolution of the mixing parameters, depending on whether the mass scheme
is hierarchical, partially degenerate or nearly degenerate.
We have listed these factors in the 
approximation of small \(\theta_{13}\) in
Tab.~\ref{tab:SuppressionEnhancementFactorsNorm}.
They can be compensated by cancellations due to a special
alignment of the phases.  For example,
an opposite CP parity of the first and second mass eigenstate,
i.e.\  $|\varphi_1-\varphi_2|=\pi$, results in a maximal suppression of
the running of the solar mixing angle, which has been pointed out
earlier in papers like
\cite{Casas:1999tg,Balaji:2000gd,Haba:2000tx,Chankowski:2001mx}.
Nevertheless, Tab.~\ref{tab:SuppressionEnhancementFactorsNorm}
allows to determine which 
angles or phases have a potential for a strong RG evolution.
Obviously, the expressions for $\Dot\delta$ are not
applicable for \(\theta_{13}=0\).  This special case will be discussed
at the end of Sec.~\ref{sec:RunningOfDelta}.
\begin{table}[!htb]
\begin{center}
 \renewcommand{\arraystretch}{1.5}
 \begin{tabular}{|c|c|c|c|c|c|}
  \hline
   & \(\Dot{\theta}_{12}\) & \(\Dot{\theta}_{13}\) & \(\Dot{\theta}_{23}\) 
   & \(\Dot{\delta}\) & \(\Dot{\varphi}_{i}\)\\
  \hline
  n.h.
  & 1 
  & \(\displaystyle\sqrt{\zeta}\)
  & \(1\)
  & \(\displaystyle\sqrt{\zeta} \, \theta_{13}^{-1}\)
  & \(\displaystyle\sqrt{\zeta}\)\\
  p.d.(n.) \rule[-3.5ex]{0ex}{1ex}
  & \(\displaystyle\frac{m_1^2}{\Delta m^2_\mathrm{sol}}\)
  & \(\displaystyle\frac{m_1}{\sqrt{\Delta m^2_\mathrm{atm}}}\)
  & \(1\)
  & \(\displaystyle\frac{m_1}{\sqrt{\Delta m^2_\mathrm{atm}}} \theta_{13}^{-1} +
      \frac{m_1^2}{\Delta m^2_\mathrm{sol}}\)
  & \(\displaystyle\frac{m_1^2}{\Delta m^2_\mathrm{sol}}\)\\
  \hline
  i.h.
  & \(\zeta^{-1}\)
  & \(\mathscr{O}(\theta_{13})\)
  & \(1\)
  & \(\zeta^{-1}\)
  & \(\displaystyle\zeta^{-1}\)\\
  p.d.(i.) \rule[-3.5ex]{0ex}{1ex}
  & \(\zeta^{-1}\)
  & \(\displaystyle\frac{m_3}{\sqrt{\Delta m^2_\mathrm{atm}}}\)
  & \(1\)
  & \(\displaystyle\frac{m_3}{\sqrt{\Delta m^2_\mathrm{atm}}}\theta_{13}^{-1}+\zeta^{-1}\)
  & \(\displaystyle\zeta^{-1}\)\\
  \hline
  d. \rule[-3.0ex]{0ex}{7ex}
  & \(\displaystyle\frac{m^2}{\Delta m^2_\mathrm{sol}}\)
  & \(\displaystyle\frac{m^2}{\Delta m^2_\mathrm{atm}}\)
  & \(\displaystyle\frac{m^2}{\Delta m^2_\mathrm{atm}}\)
  & \(\displaystyle\frac{m^2}{\Delta m^2_\mathrm{atm}}\theta_{13}^{-1}
  +\displaystyle\frac{m^2}{\Delta m^2_\mathrm{sol}}\)
  & \(\displaystyle\frac{m^2}{\Delta m^2_\mathrm{sol}}\)\\
  \hline
 \end{tabular}
\end{center}
\vspace*{-4mm}
\caption{Generic enhancement and suppression factors for the RG evolution of
the mixing parameters.
A `1' indicates that there is no generic enhancement or suppression.
`n.h.' and `p.d.(n.)' denote the hierarchical and partially
degenerate mass spectrum in the case of a normal hierarchy, i.e.\ 
\(m_1^2\ll \Delta m_\mathrm{sol}^2\) or
\(\Delta m^2_\mathrm{sol}\ll m_1^2 \lesssim \Delta m_\mathrm{atm}^2\).
`i.h.' and `p.d.(i.)' denote the analogous spectra in the inverted case,
i.e.\  \(m_3^2 \ll \Delta m_\mathrm{sol}^2\) or
\(\Delta m^2_\mathrm{sol} \ll m_3^2 \lesssim \Delta m_\mathrm{atm}^2\).
Finally, `d.' means nearly degenerate masses,
\(\Delta m^2_\mathrm{atm} \ll m_1^2 \sim m_2^2 \sim m_3^2 \sim m^2\).
}
\label{tab:SuppressionEnhancementFactorsNorm}
\end{table}

Let us consider some numerical values in order to estimate the size of
RG effects.  The SM $\tau$ Yukawa coupling is 
$y_\tau^\mathrm{SM}=\frac{\sqrt{2}}{v} m_\tau \approx 0.01$.
Thus, the typical factor in the formulae for the mixing angles and
phases amounts to
\begin{equation}
        \frac{3 y_\tau^2}{64\pi^2} \approx 0.5 \cdot 10^{-6} \;.
\end{equation}
In the MSSM it changes to
\begin{equation}
        \frac{y_\tau^2}{32\pi^2} \approx 0.3 \cdot 10^{-6}
         \left( 1+\tan^2\beta \right) \;.
\end{equation}
If the running was purely logarithmic, it would yield a factor of
\begin{equation}
        \ln \frac{M_1}{M_Z} \approx \ln \frac{10^{13}}{10^2} \approx 25
\end{equation}
for $M_1=10^{13}\,\mathrm{GeV}$.
If we assume that the solar and atmospheric angle are large and that the
phases do not cause excessive cancellations, then multiplying the above
two contributions with the enhancement factor $\Gamma_\mathrm{enh}$ from
Tab.~\ref{tab:SuppressionEnhancementFactorsNorm}
yields a rough estimate for
the change of the angles and phases due to the RG evolution,
\begin{equation} \label{eq:NaiveRGChange}
        \Delta_\mathrm{RG} \sim 
        10^{-5} \left( 1+\tan^2\beta \right) \Gamma_\mathrm{enh} \;.
\end{equation}
Of course the factor $1+\tan^2\beta$ has to be omitted in the SM.  It is
immediately clear that even in the MSSM with very large $\tan\beta$ no
significant change occurs if the enhancement factor is 1 or less --
except maybe for $\theta_{13}$, where even a change by $1^\circ$ could be
interesting.
However, for quasi-degenerate neutrinos large enhancement factors are
possible.  As an example, let us estimate the size of the absolute
neutrino mass scale (the `amount of degeneracy') needed for a sizable
RG change of $\theta_{12}$, say $0.1 \approx 6^\circ$.  In the SM, this
requires $\Gamma_\mathrm{enh} \sim 10^4$, corresponding to a neutrino
mass of the order of \(1\,\mathrm{eV}\), which is excluded by WMAP and double beta
decay experiments.  On the other hand, in the MSSM this mass scale can
easily be lowered to about \(0.1\,\mathrm{eV}\) with $\tan\beta$ as small as 8.

\subsection{Discussion and Comparison with Numerical Results}

We now study in detail the running of the mixing angles and masses, in
particular the influence of the phases.
The RG evolution of the phases will be studied separately in Sec.~\ref{sec:RunningPhases}.
We solve the RGEs for the neutrino mass operator and for the other
parameters numerically and compare the results with those obtained from
the analytical formulae of Sec.~\ref{sec:AnalyticFormulae}.
For the numerics we follow the `run and diagonalize' procedure, i.e.\ we
first compute the running of the mass matrix and then extract the
evolving mass eigenvalues and mixing parameters.  The algorithm used for
this is described in App.~\ref{app:MixingParameters}.
As an example, we consider the MSSM with \(\tan\beta=50\), a normal mass
hierarchy for the neutrinos, $m_1=0.1\,\mathrm{eV}$ for the mass of the
lightest neutrino, and a mass of about $120\,\mathrm{GeV}$ for the light
Higgs.  These boundary conditions are given at the electroweak scale, i.e.\  we
calculate the evolution from low to high energies.
Below the SUSY-breaking scale, which we take to be $1.5\,\mathrm{TeV}$, we
assume the SM to be valid as an effective theory and use the
corresponding RGEs.  Above, we apply the ones of the MSSM.

\subsubsection{RG Evolution of $\boldsymbol{\theta_{12}}$}
\label{sec:RGevolutionTheta12}

From Tab.~\ref{tab:SuppressionEnhancementFactorsNorm}, we see that  
the solar angle $\theta_{12}$ generically has the strongest 
RG effects among the mixing angles.  The reason for this is the
smallness of the solar
mass squared difference associated with it, in particular compared to
the atmospheric one, which leads to an enhanced running for quasi-degenerate
neutrinos and for the case of an inverted mass hierarchy. 
Furthermore, it is known that in the MSSM the solar angle always increases when 
running down from $M_1$ for $\theta_{13}=0$ \cite{Miura:2002nz}.  
This is confirmed by our formula \eqref{eq:AnalyticApproxT12}.
From the term 
$| {m_1}\, e^{\I \varphi_1} + {m_2}\, e^{\I  \varphi_2}|^2$ in 
Eq.~\eqref{eq:AnalyticApproxT12}, we see that a non-zero value of the
difference $|\varphi_1-\varphi_2|$ of the Majorana phases damps the RG
evolution.
The damping becomes maximal if this difference equals $\pi$, which
corresponds to an opposite CP parity of the mass eigenstates $m_1$ and
$m_2$. 
This is in agreement with earlier studies, e.g.\  
\cite{Casas:1999tg,Balaji:2000gd,Haba:2000tx,Chankowski:2001mx}.

\begin{figure}[htb]
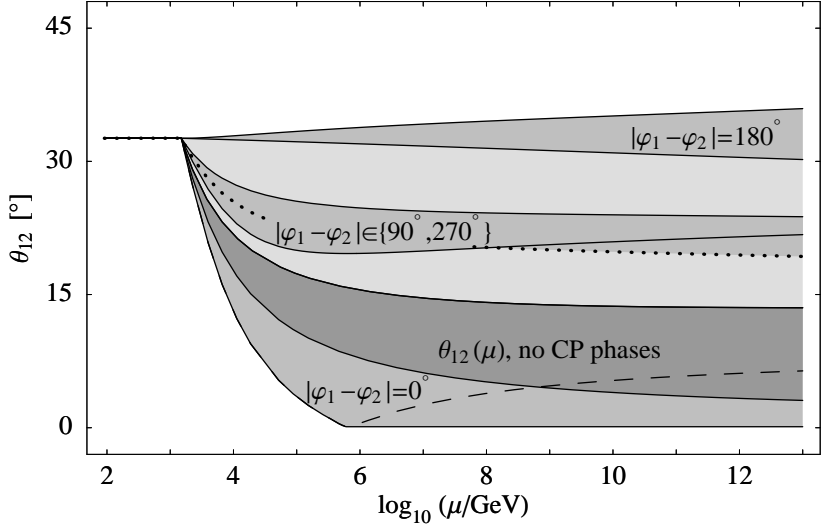

 \begin{center}
        {$\CenterEps{Theta12Running}$}
 \vspace*{-3mm}
 \end{center}
 \caption{\label{fig:PhasesMSSMtb50_t12} RG evolution of $\theta_{12}$ in the 
 MSSM with \(\tan\beta=50\), a normal mass hierarchy and $m_1=0.1\,\mathrm{eV}$.
  The dark-gray region shows the evolution with best-fit values for the neutrino
  parameters, $\theta_{13}\in [0^\circ,9^\circ]$ and all CP phases equal 
  to zero. 
  The medium-gray regions show the evolution for 
  $|\varphi_1-\varphi_2|=0^\circ$, $|\varphi_1-\varphi_2|
  \in \{90^\circ,270^\circ\}$ and $|\varphi_1-\varphi_2|=180^\circ$.
  They emerge from varying 
   $\theta_{13}\in [0^\circ,9^\circ]$ and $\delta\in 
  \{0^\circ,90^\circ,180^\circ,270^\circ\}$. 
  The light-gray regions can be
  reached by choosing specific values for the CP phases
  different from the ones listed above. The
  dashed line shows the RG evolution with $|\varphi_1-\varphi_2|=0$, 
  $\theta_{13}=9^\circ$ and $\delta=180^\circ$. Note that 
  for the numerics we use the convention where $\theta_{12}$ is
  restricted to the interval $[0^\circ,45^\circ]$, so that the angle
  increases again after reaching 0.  The dotted line shows the
  evolution with $|\varphi_1-\varphi_2|=90^\circ$ and 
  $\theta_{13}=0^\circ$.
  }
\end{figure}

Let us now compare the analytical approximation for $\Dot{\theta}_{12}$ of 
Eq.~\eqref{eq:AnalyticApproxT12} with the numerical solution for the
running in the case of nearly degenerate masses, which is shown in 
Fig.~\ref{fig:PhasesMSSMtb50_t12} in detail.
The dark-gray region shows the evolution with LMA best-fit values for the neutrino
parameters, $\theta_{13}$ varying in the interval $[0^\circ,9^\circ]$
and all CP phases equal to zero. 
The medium-gray regions show the evolution for   
$|\varphi_1-\varphi_2|\in \{0^\circ,90^\circ,180^\circ,270^\circ\}$,
$\theta_{13}\in [0^\circ,9^\circ]$ and 
$\delta\in \{0^\circ,90^\circ,180^\circ,270^\circ\}$, 
confirming the expectation of the damping influence of \(\varphi_1\) and
\(\varphi_2\).  The flat line at low energy stems from the SM running
below $M_\mathrm{SUSY}$, which is negligible as we have seen earlier.
Note that the numerics never yield negative values of $\theta_{12}$ due
to the algorithm used for extracting the mixing parameters from the MNS
matrix, which guarantees $0 \leq \theta_{12} \leq 45^\circ$ (see
App.~\ref{sec:LeptMixingMatrix} for further details).

As can be seen from the relatively broad dark-gray band in the figure,
the $\mathscr{O}(\theta_{13})$-term in the RGE is quite important here.
The dominant part of this term is
\begin{eqnarray}
 \Upsilon
& = & 
 \frac{C y_\tau^2}{32\pi^2}
 \frac{m_2+m_1}{m_2-m_1}
 \, \sin 2\theta_{23} \,
 \cos\frac{\varphi_1-\varphi_2}{2} \times
\nonumber\\*
&& \quad {} \times
 \Bigl(
  \cos 2\theta_{12} \,\cos\delta\,\cos\frac{\varphi_1-\varphi_2}{2}
  +\sin\delta\,\sin\frac{\varphi_1-\varphi_2}{2}
 \Bigr) \cdot \theta_{13}
 \;.\label{eq:AnalyticApproxT12OT13}
\end{eqnarray}
Clearly, the RG evolution of \(\theta_{12}\) is independent of the Dirac
phase \(\delta\) only in the approximation \(\theta_{13}=0\).
The largest running, where \(\theta_{12}\) can even become zero, 
occurs for  \(\theta_{13}\) as large as possible ($9^\circ$), 
$\delta=\pi$ and $\varphi_1 - \varphi_2 = 0$.  In this case
the leading and the next-to-leading term add up constructively.
It is also interesting to observe that due to $\mathscr{O}(\theta_{13})$ effects
\(\theta_{12}\) can run to slightly larger values. The damping due to the 
Majorana phases is maximal in this case, which almost
eliminates the leading term. Then, all the running comes from the
next-to-leading term \eqref{eq:AnalyticApproxT12OT13}.

In the inverted scheme, \(m_1\gg m_2-m_1\) always holds, so that large RG effects 
are generic, i.e.\  always present except for the case of cancellations due to
Majorana phases. For a normal mass hierarchy with a small $m_1$, the
running of the solar mixing is of course rather insignificant.

Finally, we would like to emphasize that it is not appropriate to assume
the right-hand sides of Eq.~\eqref{eq:AnalyticApproxT12} and
Eq.~\eqref{eq:AnalyticApproxT12OT13} to be constant in order to interpolate
$\theta_{12}$ up to a high energy scale, since non-linear effects
especially from the running of $\sin 2\theta_{12}$ and $\Delta
m^2_\mathrm{sol}$ cannot be neglected here.  This is easily seen from
the curved lines in Fig.~\ref{fig:PhasesMSSMtb50_t12}.

\subsubsection{RG Evolution of $\boldsymbol{\theta_{13}}$}
\label{sec:RunningTheta13}

\begin{figure}[!htb]
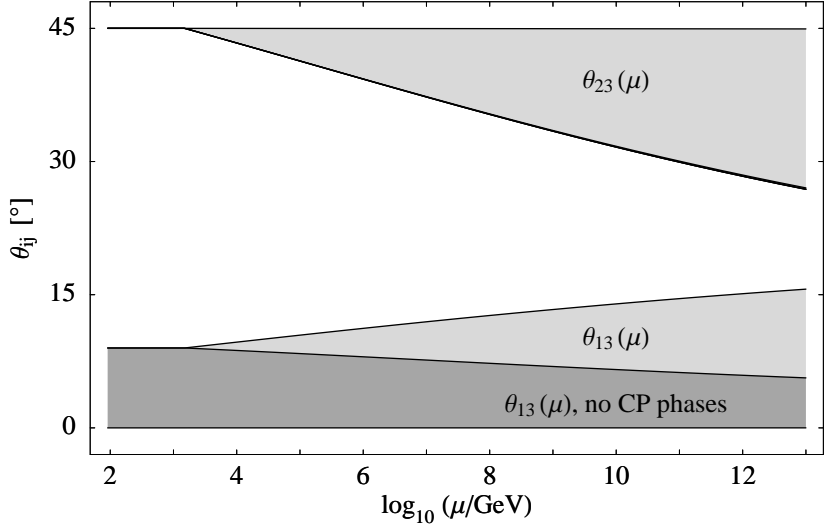

 \begin{center}
        {$\CenterEps{Theta13and23Running}$}
 \vspace*{-3mm}
 \end{center}
 \caption{\label{fig:PhasesMSSMtb50_t13} RG evolution of $\theta_{13}$ and 
 $\theta_{23}$ in the MSSM with \(\tan\beta=50\), 
 a normal mass hierarchy and $m_1=0.1\,\mathrm{eV}$.
  The dark-gray region shows the evolution with best-fit values for the neutrino
  parameters, $\theta_{13}\in [0^\circ,9^\circ]$ and all CP phases equal 
  to zero. For the $\theta_{23}$ case, we just obtain a thick gray line at the
  bottom of the gray region. 
  The light-gray regions show the evolution, which is possible, if
  arbitrary CP phases are allowed. 
  }
\end{figure}

The analytical approximation for $\Dot{\theta}_{13}$ is given in  
Eq.~\eqref{eq:AnalyticApproxT13}. 
As already pointed out, in order to apply it to
the case $\theta_{13}=0$, where $\delta$ is undefined, the analytic 
continuation of the latter has to be inserted. It will  
be given in Eq.~\eqref{eq:DeltaZeroT13} in section \ref{sec:RunningOfDelta}, 
where the phases are treated in detail.
The comparison with the numerical results in
Fig.~\ref{fig:PhasesMSSMtb50_t13} shows that above $M_\mathrm{SUSY}$ the
angle runs linearly on a logarithmic scale to a good approximation.
Thus, using Eq.~\eqref{eq:Theta13Dot} with a constant right-hand side
yields pretty accurate results.
With $\varphi_1\neq\varphi_2$, significant RG
effects can be expected for nearly degenerate masses.  This is confirmed
by the light-gray region in Fig.~\ref{fig:PhasesMSSMtb50_t13}.  

The fastest running occurs if $\varphi_1-\varphi_2=\pi$
and $\varphi_1-\delta \in \{0,\pi\}$, so that the
terms proportional to $m_1$ and $m_2$ in the RGE are maximal and add up.  
Interestingly, cancellations between the first two terms in the second
line of Eq.~\eqref{eq:Theta13Dot} appear for $\varphi_1=\varphi_2$,
in particular if all phases are zero. If so, the leading contribution to
the evolution of $\theta_{13}$ is suppressed by an additional factor of
$\zeta$.  This suppression is in agreement with earlier studies, for
instance \cite{Balaji:2000gd,Bhattacharyya:2002aq}, where it was
discussed for the CP-conserving case \(\varphi_1=\varphi_2=\pi\), which
implies an opposite CP parity of $m_3$ compared to the other two mass
eigenvalues. 
Such cancellations cannot occur for a strong normal mass hierarchy, since then the
evolution is dominated by the term proportional to $m_2$ in 
Eq.~\eqref{eq:AnalyticApproxT13}.  

Besides, $\theta_{13}$ runs towards smaller values in the MSSM
with zero phases and a normal hierarchy, because $m_1<m_2$, so that the second line
of the RGE is negative.  This yields the dark-gray region in
Fig.~\ref{fig:PhasesMSSMtb50_t13}.\footnote{The relatively 
large slope of its upper boundary is due to the $\mathscr{O}(\theta_{13})$ 
contribution to the RGE.} 
As $\theta_{13}$ can always be made positive by a suitable redefinition
of parameters, the sign of $\Dot\theta_{13}$ is irrelevant for
$\theta_{13}=0$.

For an inverted hierarchy, the situation is reversed, since
$\Delta m^2_\mathrm{atm}$ is negative then.  For a small $m_3$, the
running is highly suppressed in this case, because the leading term is
proportional to $m_3$.  Then the dominant contribution comes from the
$\mathscr{O}(\theta_{13})$-term unless $\theta_{13}$ is very small as
well.

Future experiments will probably be able to probe $\sin^2 2\theta_{13}$
down to $10^{-4}$, corresponding to
$\theta_{13} \sim 5\cdot10^{-3} \sim 0.3^\circ$.
Consequently, even RG changes of this order of magnitude could be
important, since a low-energy value smaller than the RG change would
appear unnatural.  This will be discussed in more detail in
Sec.~\ref{sec:ConstraintsFromRG}.

\subsubsection{RG Evolution of $\boldsymbol{\theta_{23}}$}
The analytical RGE for $\Dot{\theta}_{23}$ can be found in
Eq.~\eqref{eq:AnalyticApproxT23}. 
Again, the comparison with the numerical results (see 
Fig.~\ref{fig:PhasesMSSMtb50_t13})  
shows that to a good approximation the angle 
runs linearly on a logarithmic scale above $M_\mathrm{SUSY}$. 
The sign of $\Delta m^2_\mathrm{atm}$ is very important here.  For a
normal mass spectrum, the leading term is always negative in the MSSM,
so that $\theta_{23}$ decreases with increasing energy, while for an
inverse spectrum the situation is exactly reversed, so that
$\theta_{23}$ becomes larger than $45^\circ$ if one starts with the LMA
best-fit value at low energy.

From Eq.~\eqref{eq:AnalyticApproxT23} we expect that switching on the
phases $\varphi_1$ and $\varphi_2$ always reduces the running of
\(\theta_{23}\) for nearly degenerate masses.  This is confirmed by the
light-gray region in Fig.~\ref{fig:PhasesMSSMtb50_t13}.
The damping is much less severe for a hierarchical mass spectrum, since
either $m_1$ and $m_2$ or $m_3$ are very small then.  However, in these
cases the running is generally expected to be rather insignificant,
since according to Tab.~\ref{tab:SuppressionEnhancementFactorsNorm} the
enhancement factor is only 1.

\subsubsection{RG Evolution of the Neutrino Mass Eigenvalues}
\label{sec:RunningOfMasses}
The running of the mass eigenvalues is significant even in the SM or for
strongly hierarchical neutrino masses due to the factor $\alpha$ in the
RGEs \eqref{eq:EvolutionOfMassEigenvalues}.
Clearly, the evolution is not directly 
dependent on the Majorana phases \cite{Casas:1999tg}.  This can be understood from 
Eqs.~\eqref{eq:EvolutionOfEffectiveMasses} and \eqref{eq:PiiPrimeReal}, which
show that only the moduli of the elements of the MNS matrix enter into
$\Dot{m}_i$.  Besides, \(\Dot{m}_3\) does not depend on \(\delta\),
since only the moduli of the elements of the third column of the MNS
matrix are relevant in this case.  Of course, there is an indirect
dependence on the phases, as these influence the running of the mixing
angles.

\begin{figure}[!htb]
\begin{center}
  $\CenterEps[0.9]{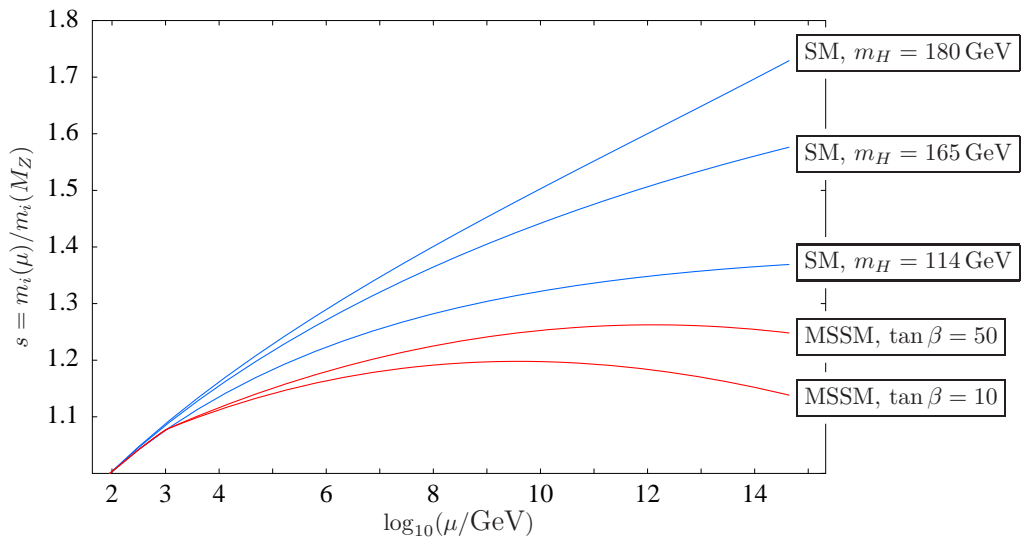}$
\end{center}
 \vspace*{-3mm}
 \caption{
 Scaling of the masses under the renormalization group  
 in the SM and MSSM. 
 The mixing parameters are chosen to be the LMA best-fit values (cf.\
 Tab.~\ref{tab:ExpData}), but they influence the running only marginally.
 We further used a SUSY-breaking scale \(M_\mathrm{SUSY}=1\,\mathrm{TeV}\). 
 The upper curves show the evolution in the SM for \(m_H=114\,\mathrm{GeV}\),
 \(m_H=165\,\mathrm{GeV}\) and \(m_H=180\,\mathrm{GeV}\), 
 the lower ones correspond to the MSSM for \(\tan\beta=10\) 
 and \(\tan\beta=50\) with \(m_H=120\,\mathrm{GeV}\).
 These plots apply for all mass eigenvalues, except for large \(\tan\beta\) 
 in the MSSM where the scaling of \(m_3\) is shown (using zero phases). Note also that a 
 different SUSY-breaking scale changes the scaling factor in the MSSM.
 } 
 \label{fig:ScalingOfMasses}
\end{figure}

Apart from the MSSM with large $\tan\beta$, the running of the mass eigenvalues
is virtually independent of the mixing parameters, since $\alpha$ is 
usually much larger than $y_\tau^2$.  
In the SM, the Higgs mass influences the running via the
self-coupling $\lambda$ -- the heavier the Higgs, the larger the RG
effects.
Thus, except for large \(\tan \beta\) in the MSSM, the running is 
given by a common scaling of the mass eigenvalues \cite{Chankowski:2001mx},
which is obtained by neglecting \(y_\tau\) and integrating 
Eq.~\eqref{eq:EvolutionOfMassEigenvalues},
\begin{equation}
 m_i(t)
 \,\approx\,
 \exp\left[\frac{1}{16\pi^2} \int_{t_0}^t\D\tau\,\alpha(\tau)\right]\,m_i(t_0)
 \,=:\, s(t,t_0)\, m_i(t_0)
 \;.
\end{equation}
We plot \(s\) in the SM and in the MSSM for various parameter combinations 
in Fig.~\ref{fig:ScalingOfMasses}. 
The three SM curves correspond to different Higgs masses in the
current experimentally allowed region at 95\% confidence level, 
\(114\,\mathrm{GeV}\lesssim m_H \lesssim200\,\mathrm{GeV}\)
\cite{Higgsbound}.  \(m_H=180\,\mathrm{GeV}\)
is the value for which the self-coupling \(\lambda\) stays
perturbative up to \(10^{16}\,\mathrm{GeV}\), i.e.\ $\lambda\lesssim1$,
and \(m_H=165\,\mathrm{GeV}\) 
is the minimal mass for which \(\lambda\) is positive up to \(10^{16}\,\mathrm{GeV}\),
so that the vacuum is stable in this region 
(see e.g.~\cite{Cabibbo:1979ay,Lindner:1986uk}).\footnote{In some models
(see, e.g.\  \cite{Antusch:2002xh} for a viable model) \(\lambda\) can be
larger, in particular if \(M_1\ll 10^{16}\,\mathrm{GeV}\). A negative
value of \(\lambda\) at high energy implies a metastable vacuum.} 
In the MSSM, we choose \(m_H=120\,\mathrm{GeV}\) for the light Higgs
mass, since the allowed range is further restricted by the upper limit
at about \(130\,\mathrm{GeV}\) here, and since it influences the
evolution of the RG scaling only marginally as long as
\(M_\mathrm{SUSY}\) and \(M_Z\) differ only by a few orders of magnitude.
Moreover, further uncertainties
due to threshold corrections and the unknown value of the SUSY-breaking
scale can be equally important as the one due to the unknown Higgs mass.
The RG enhancement of the masses is smallest if \(\tan\beta\approx 10\).

\begin{figure}[htb]
\begin{center}
    \subfigure[Normal mass hierarchy\label{fig:RunningMassesr}]{
                 $\CenterEps[0.9]{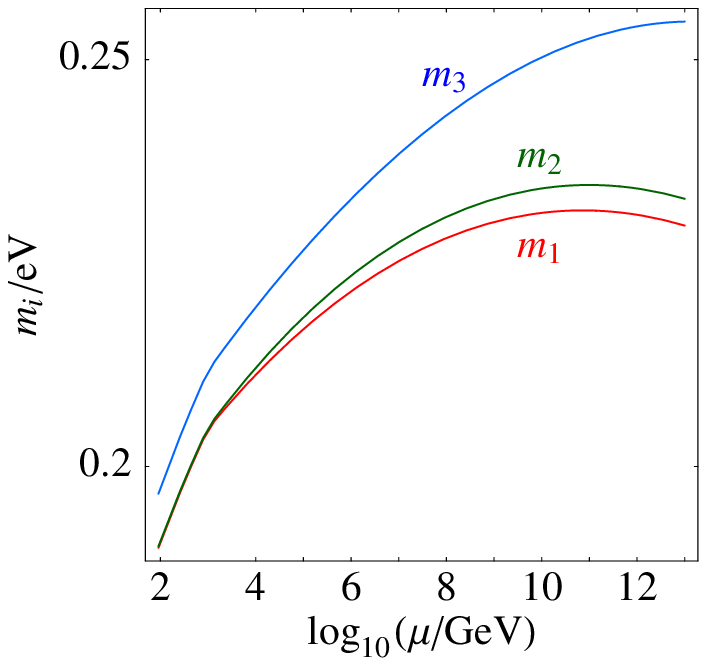}$} \hfil
\subfigure[Inverted mass hierarchy\label{fig:RunningMassesi}]{
                 $\CenterEps[0.9]{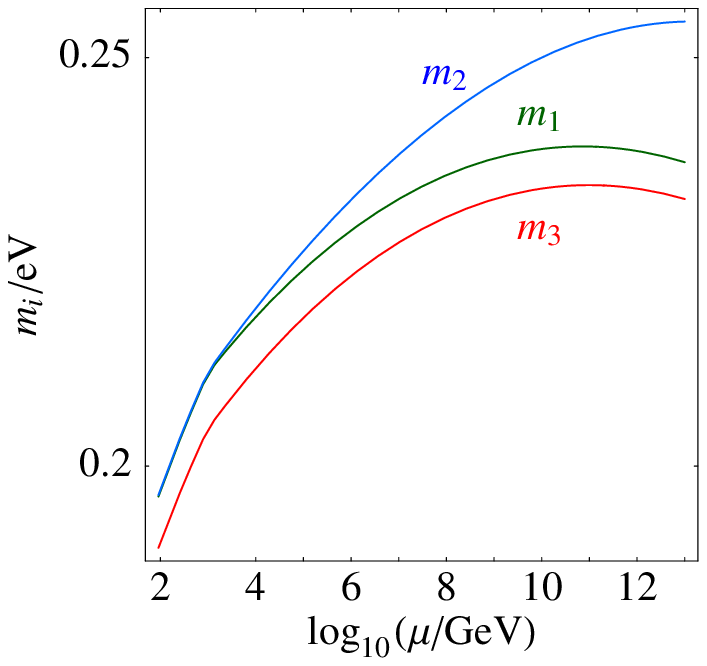}$}               
\end{center}
 \vspace*{-5mm}
 \caption{
 Running of the light neutrino masses
 for a normal and an inverted mass hierarchy and 
 \(m_\mathrm{min}=0.19\,\mathrm{eV}\) in the MSSM with \(\tan\beta=50\)
 and \(M_\mathrm{SUSY}=1\,\mathrm{TeV}\).
 The mixing parameters are chosen to be the LMA best-fit values.
 The phases are zero in this example.
 In the inverted case, \(\Delta m^2_\mathrm{sol}\) becomes
 greater than \(|\Delta m^2_\mathrm{atm}|\).
 } 
 \label{fig:RunningMasses}
\end{figure}

As already mentioned, substantial deviations from the common scaling
arise in the MSSM for large \(\tan\beta\). There is a plethora of
effects which can be understood with the aid of 
\eqref{eq:EvolutionOfMassEigenvalues} and \eqref{eq:RunningDeltaM2s}. 
In order to give an interesting example, we show the evolution of the 
mass eigenvalues for \(m_\mathrm{min}=0.19\,\mathrm{eV}\) (where
\(m_\mathrm{min}=\min\{m_1,m_2,m_3\}\)) in the MSSM with \(\tan\beta=50\)
in Fig.~\ref{fig:RunningMasses}. A particular interesting effect is 
that for an inverted mass spectrum the property
\(|\Delta m^2_\mathrm{atm}|>\Delta m^2_\mathrm{sol}\) possibly does not
survive the RG evolution. In other words, what looks like a normal
mass hierarchy at high energies turns out to become an inverted hierarchy
at low energies (cf.\ Fig.~\ref{fig:RunningMassesi}).
From the dependence on the \(y_\tau^2\) terms (cf.\ Eqs.\
\eqref{eq:Fi} and \eqref{eq:FsolAndFatm}), we find
that this effect can disappear if \(\delta\) is large.

\subsubsection{RG Evolution of $\boldsymbol{\Delta m_\mathrm{sol}^2}$}
\label{sec:RunningDm2sol}
\begin{figure}[!htb]
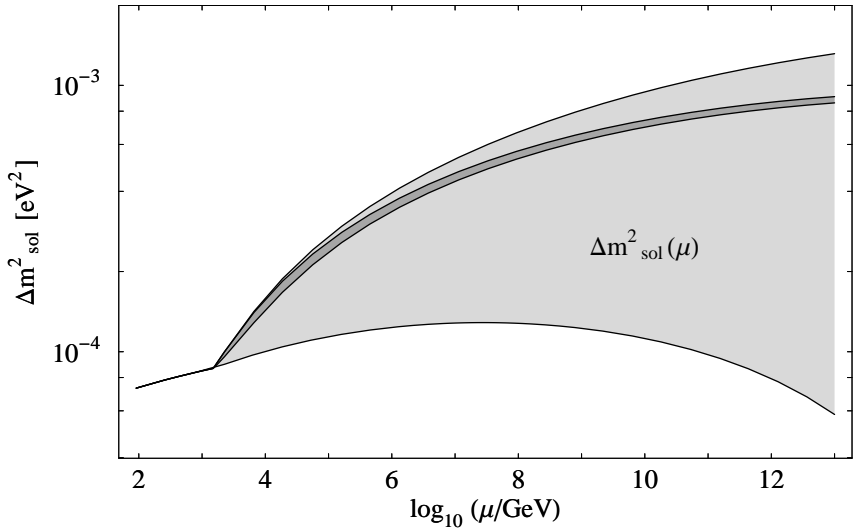

 \vspace*{3mm}
 \begin{center}
        $\CenterEps{DeltaM2solRunning}$
 \end{center}
 \vspace*{-1mm}
 \caption{
 RG evolution of $\Delta m_\mathrm{sol}^2$ 
 in the MSSM with \(\tan\beta=50\), 
 a normal mass hierarchy and $m_1=0.1\,\mathrm{eV}$.
 The dark-gray region shows the evolution with LMA best-fit values for the neutrino
  parameters, $\theta_{13}\in [0^\circ,9^\circ]$ and all CP phases equal 
  to zero.  
  The light-gray regions show the evolution, which is possible, if
  arbitrary CP phases are allowed. 
  }
  \label{fig:PhasesMSSMtb50_m2Dsol}
\end{figure}

The RGE for the solar mass squared difference is given in
Eq.~\eqref{eq:Dm2solDot}.
In the SM and the MSSM with small $\tan\beta$, the running is due to the
common scaling of the masses described in the previous section and thus
virtually independent of the mixing parameters.
For large $\tan\beta$ and nearly degenerate masses, the influence of CP
phases, in particular the Dirac phase, is crucial. The numerical example
in Fig.~\ref{fig:PhasesMSSMtb50_m2Dsol} confirms this expectation and
furthermore shows that \(\Delta m_\mathrm{sol}^2\) runs dramatically.
On the one hand, it can grow by more than an order of magnitude.  As we
have seen in Fig.~\ref{fig:RunningMasses}, \(\Delta m_\mathrm{sol}^2\)
can even get larger than \(|\Delta m_\mathrm{atm}^2|\).
On the other hand, it can run to $0$ at energy scales slightly beyond
the maximum of $10^{13}\,\mathrm{GeV}$ shown in the figure.
For large \(\tan\beta\), \(\Delta m_\mathrm{sol}^2\ll m_1^2\) and not
too small \(\theta_{13}\), the first term in $F_\mathrm{sol}$ is
essential for understanding these effects, since it is proportional to
the sum of the masses squared rather than the difference.  For 
$\delta=\pi$ and $\theta_{13}$ near the CHOOZ bound, its sign is
negative and its absolute value maximal, which causes
the evolution of $\Delta m_\mathrm{sol}^2$ towards zero.
For $\delta=0$, the sign becomes positive, so that the running towards
larger values is enhanced, which explains the upper boundary of the
light-gray region in Fig.~\ref{fig:PhasesMSSMtb50_m2Dsol}.

\subsubsection{RG Evolution of $\boldsymbol{\Delta m_\mathrm{atm}^2}$}
\label{sec:RunningDm2atm}
\begin{figure}[htb]
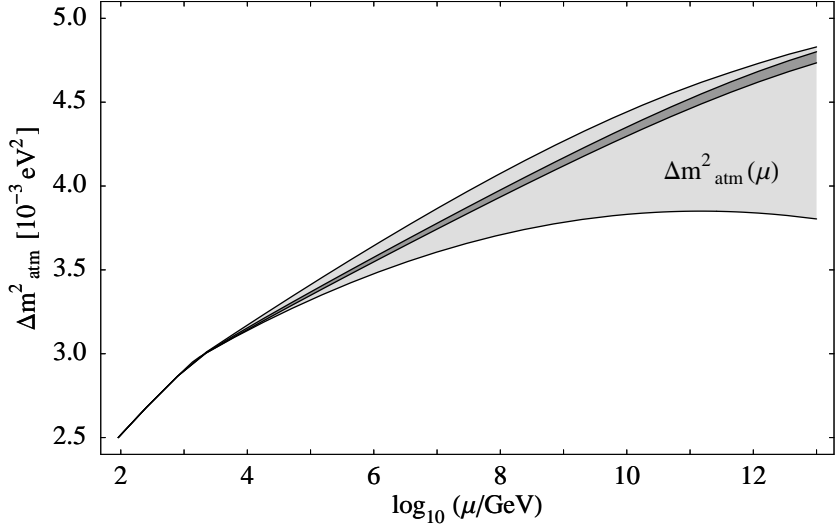

 \begin{center}
        $\CenterEps{DeltaM2atmRunning}$
 \vspace*{-3mm}
 \end{center}
 \caption{\label{fig:PhasesMSSMtb50_m2D} 
 RG evolution of $\Delta m_\mathrm{atm}^2$
 in the MSSM with
 the same input parameters as in Fig.~\ref{fig:PhasesMSSMtb50_m2Dsol}.
 }
  \label{fig:PhasesMSSMtb50_m2Datm}
\end{figure}

From the numerical example in Fig.~\ref{fig:PhasesMSSMtb50_m2Datm}, we
see that $\Delta m_\mathrm{atm}^2$ can be damped by the phases, but not
significantly enhanced.  Depending on the CP phases, 
$\Delta m_\mathrm{atm}^2$ grows by about 50\% -- 95\%.  Analogously to
above, the maximal damping is mainly due to the first term in $F_\mathrm{atm}$,
so that it occurs for large $\theta_{13}$ and $\delta=0$.  Compared to
the case of the solar mass squared difference, the influence of
\(\delta\) is generically smaller here, because 
$\Delta m_\mathrm{atm}^2/m_i^2$ is larger and because the
phase-independent terms in the RGE do not nearly cancel.

\newpage

\subsection{RG Running of the Dirac and Majorana Phases}
\label{sec:RunningPhases}

Most earlier studies of RG effects either neglected phases or
concentrated on the special case of a Majorana parity, where one or both
of the Majorana phases are $\pi$.  We have seen that they can
have a dramatic influence on the running of the masses and mixings. 
Moreover, many effects are affected by phases, e.g.\  neutrinoless double
beta decay, or require phases, e.g.\ leptogenesis.%
\footnote{Clearly, the phases relevant for leptogenesis
are those of the `right-handed' sector and therefore in general not directly 
related to the phases considered here \cite{Branco:2001pq,Pascoli:2003uh}. 
However, as the left-handed sector with its -- in principle -- observable phases
is related to the right-handed one by the see-saw relation, it is reasonable to
assume that non-vanishing right-handed phases imply non-zero \(\delta\),
\(\varphi_1\) and/or \(\varphi_2\). An explicit relation which supports this
point of view is specified in, e.g., \cite{Frampton:2002qc}.}

Of course, if the phases are given at some scale, they also change due
to the RG evolution.
We now discuss the running of the phases themselves and give numerical examples.
In general, a significant evolution of the phases is expected for
nearly degenerate and inverted hierarchical mass patterns, since the RGEs
\eqref{eq:DeltaPrimeWithNonZeroDelta}--\eqref{eq:Phi1Dot} contain the
ratios 
$m_1 m_2 / \Delta m^2_\mathrm{sol}$.

\subsubsection{RG Evolution of the Dirac Phase}
\label{sec:RunningOfDelta}
\begin{figure}[!htb]
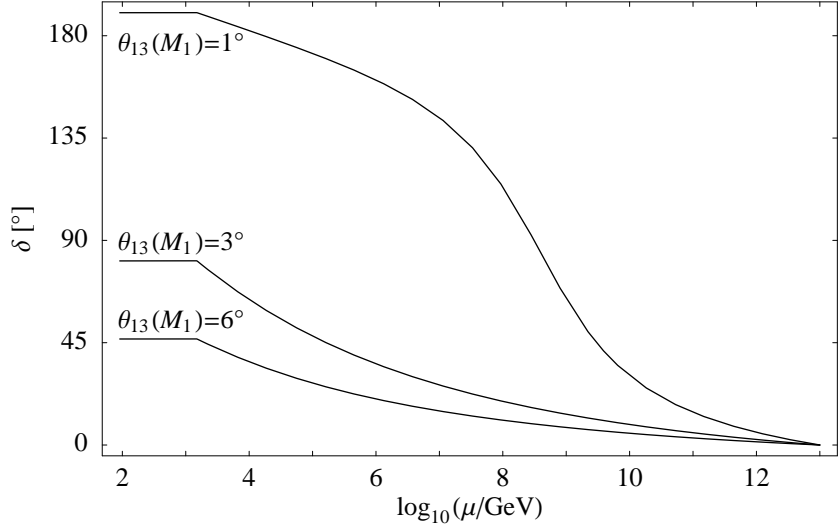

 \begin{center}
 \CenterEps{PhaseEvolution12}
 \end{center}
 \caption{Radiative generation of a Dirac phase in the MSSM with
  $\tan\beta=30$ and a normal hierarchy.  Here the running is from high
  to low energy, i.e.\  the boundary conditions are given at the see-saw
  scale.  $\delta$ is zero there but large at $M_Z$.  The other starting
  values are
  $\theta_{12} = 18^\circ$,
  $\theta_{13} \in \{1^\circ, 3^\circ, 6^\circ\}$,
  $\theta_{23} = 34^\circ$,
  $m_1 = 0.17\,\mathrm{eV}$,
  $\Delta m^2_\mathrm{atm} = 3.8 \cdot 10^{-3}\,\mathrm{eV}^2$, 
  $\Delta m^2_\mathrm{sol} = 5.7 \cdot 10^{-4}\,\mathrm{eV}^2$,
  $\varphi_1 = 16^\circ$, $\varphi_2 = 140^\circ$.
 }
 \label{fig:DiracPhaseEvolution5}
\end{figure}

The running of the Dirac phase $\delta$ is given by
Eq.~\eqref{eq:DeltaPrimeWithNonZeroDelta} for $y_e=y_\mu=0$.
An interesting possibility is the radiative generation of a Dirac phase
by Majorana phases \cite{Casas:1999tg}: A non-zero $\delta$ is produced by RG effects,
since some of the terms in the RGE \eqref{eq:DeltaPrimeWithNonZeroDelta}
do not vanish for $\delta\rightarrow0$.  Fig.~\ref{fig:DiracPhaseEvolution5} shows
an example.  The most important term in this context is the first one in
$\delta^{(0)}$.  As it is proportional to
$\sin(\varphi_1-\varphi_2)$, the effect is suppressed for
$\varphi_1=\varphi_2$.  For small but non-zero values of $\theta_{13}$,
the term involving $\delta^{(-1)}$ also contributes
significantly because of the factor $\theta_{13}^{-1}$.  For
$\varphi_1=\varphi_2$, this contribution is suppressed as well, since
the parts proportional to $m_1$ and $m_2$, respectively, nearly cancel.

In the case of an inverted hierarchy with $\tan\beta$ varying between 30
and 50, Dirac phases of about $15^\circ$ to $30^\circ$ can be generated.
Now the term involving $\delta^{(-1)}$ receives an additional
suppression from the small value of $m_3$, so that the subleading
effects described above become unimportant.  Hence, the running of
$\delta$ is independent of $\theta_{13}$ and depends only on the
difference of the Majorana phases to a very good approximation.

Before we turn to the evolution of the Majorana phases, let us discuss
some further properties of the RGE for $\delta$ that are also valid
beyond the special case of a radiative generation of this phase.  To
start with, the most important term in $\Dot\delta$ depends only on the
difference of the Majorana phases.  Consequently, the evolution is
expected to stay roughly the same if both phases change by the same
value.  A comparison with numerical results shows that this is true only
to a first approximation.  If one starts with $\varphi_2=0$ and
increments it step by step, the running of $\delta$ is increasingly
damped.  The main reason for this is the second term in square brackets
in $\delta^{(-1)}$  (the one proportional to $m_2$), whose sign is
opposite to that of the leading term for $\delta<\varphi_2$.  This term
grows with $\varphi_2$, while the previous one (proportional to $m_1$)
does not change much as long as $\varphi_1$ is close to $90^\circ$.
The situation can be very different for smaller values of $\theta_{13}$.
Now the initial rise of $\delta$ is enhanced, so that it can become
larger than $\varphi_2$.  Then the sign of the aforementioned second
term in square brackets changes, so that it no longer damps the
evolution but amplifies it.

With a strong normal hierarchy, RG effects are usually tiny.  The
running of the Dirac phase is one of the few examples where this is not
always the case.  Due to the terms proportional to $\theta_{13}^{-1}$ in
the RGE, a significant evolution is possible for small $\theta_{13}$.
However, one has to keep in mind that a measurement of $\delta$ is very
hard in this case.

Regardless of the mass hierarchy, the limit $\theta_{13} \to 0$ is
dangerous, because in this case
the RGE \eqref{eq:DeltaPrimeWithNonZeroDelta} diverges.  However, we can
show that $\Dot{\delta}$ remains well-defined: The derivative of the MNS
matrix $U$ is given by \eqref{eq:EvolutionOfU}, $\Dot{U} = U \cdot T$,
where $U$ and $T$ are continuous.
Hence, $U_{13}(t)$ describes a continuously differentiable curve in the
complex plane.  Consequently, $\theta_{13}$ and $\delta$ are
continuously differentiable even for $\theta_{13}=0$, if $\delta$ is
extended continuously at this point.
Note that restricting the parameters to certain ranges can nevertheless
result in discontinuities.  For example, if the RG evolution causes 
$\theta_{13}$ to change its sign and if we demand
$0\leq\theta_{13}<\frac{\pi}{2}$, then there will be a kink in the
evolution of $\theta_{13}$ and $\delta$ will jump by $\pi$.
However, even in the presence of such artificial discontinuities there
must still be finite one-sided limits for $\delta$ and $\Dot{\delta}$ as
$\theta_{13}$ approaches 0.

The limit for $\delta$ is determined by the requirement that
$\Dot{\delta}$ remains finite.  Then the divergence of $\theta_{13}^{-1}$
has to be canceled by $\delta^{(-1)}$.
For $\varphi_1=\varphi_2=0$, this obviously implies $\delta=0$ or
$\delta=\pi$.  In the general case, a short calculation yields
\begin{equation}\label{eq:DeltaZeroT13}
        \cot\delta =
        \frac{ m_1 \cos\varphi_1 - \left(1+\zeta\right) m_2 \,\cos\varphi_2-
         \zeta m_3 }
        { m_1 \sin\varphi_1 - \left(1+\zeta\right) m_2 \, \sin\varphi_2 }
        \;.
\end{equation}
Due to the periodicity of $\cot$, there are two solutions differing by
$\pi$, corresponding to the different limits on the two sides of a node
of $\theta_{13}$.

\subsubsection{RG Evolution of the Majorana Phases}
\begin{figure}[!htb]
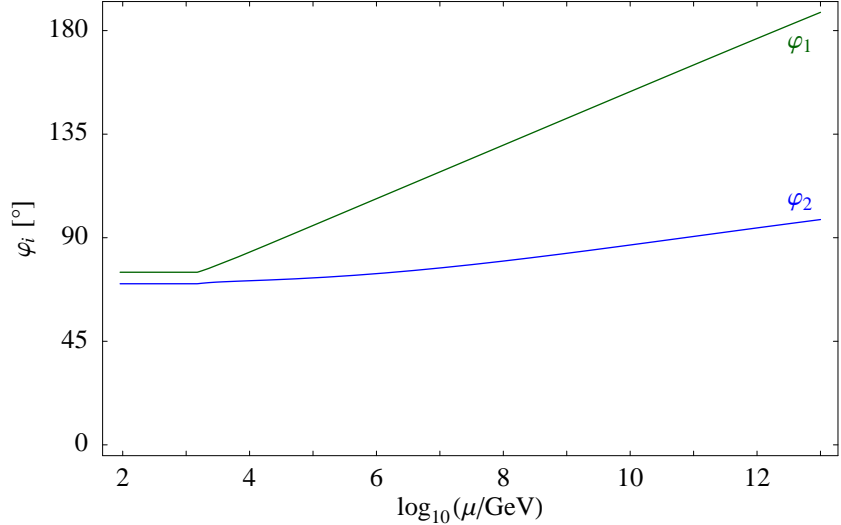

 \begin{center}
 \CenterEps{PhaseEvolution2}
 \end{center}
 \caption{Running of the Majorana phases
  in the MSSM with a normal hierarchy,
  $\tan\beta=50$, $\varphi_1=75^\circ$, $\varphi_2=70^\circ$,
  $\theta_{13}=0$, $m_1=0.15\,\mathrm{eV}$, and LMA best-fit values for the mass
  squared differences, $\theta_{12}$ and $\theta_{23}$ at $M_Z$.
  RG effects are substantial, and the difference $\varphi_1-\varphi_2$
  increases with increasing energy.
 }
 \label{fig:MajPhaseEvolution1}
\end{figure}

While the RGEs for the Majorana phases are somewhat lengthy, there is a
simple expression for the running of their difference for small
$\theta_{13}$,
\begin{equation} \label{eq:DeltaPhiEvolution}
        \Dot\varphi_1 - \Dot\varphi_2 =
        \frac{C y_\tau^2}{4\pi^2} \,
        \frac{ m_1 m_2 }{ \Delta m^2_\mathrm{sol} } \,
        \cos 2\theta_{12} \,\sin^2\theta_{23} \, \sin(\varphi_1-\varphi_2) +
        \mathscr{O}(\theta_{13}) \;.
\end{equation}
It shows that for $\theta_{13}=0$, the phases remain equal, if they are
equal at some scale.
Obviously, $\Dot\varphi_1-\Dot\varphi_2 > 0$ for $\varphi_1>\varphi_2$
and vice versa, which means that the difference between the phases tends
to increase with increasing energy.  In other words, a large difference
at the see-saw scale becomes smaller at low energy.  An example is shown
in Fig.~\ref{fig:MajPhaseEvolution1}.

If $\varphi_1-\varphi_2$ is not too small, a non-zero $\theta_{13}$
tends to damp its running.  This is due to a term in the RGE for
$\varphi_1$ whose sign is opposite to that of the leading one in
Eq.~\eqref{eq:DeltaPhiEvolution} and which is proportional to
$\sin\theta_{13} \, \cot\theta_{12}$.  This term can grow important if
$\theta_{12}$ becomes small with increasing energy.

For $\varphi_1=\varphi_2$ the evolution of the Majorana phases is
suppressed, since the leading
terms in the RGEs \eqref{eq:Phi1Dot} and \eqref{eq:Phi2Dot} are zero
then.  However, for larger $\tan\beta$ RG effects are still important.
Non-linear effects caused by the
decrease of the solar and atmospheric mixing angles are essential here,
as the initial slope of the curves is extremely small due to the
suppression by $\sin\theta_{13}$ and $\cos 2\theta_{23}$.  For
$\theta_{13}=5^\circ$, the second line in the RGE and the terms
proportional to $\sin\theta_{13}$ are about equally important for the
running of $\varphi_1$.  The evolution of $\varphi_2$ is virtually
independent of $\theta_{13}$, since the respective terms are not 
multiplied by $\cot\theta_{12}$, which again can become large as the energy
increases because of the diminishing $\theta_{12}$, but by
$\tan\theta_{12}$, which remains smaller than 1.

In principle, it is also possible to generate Majorana phases
radiatively, if the CP phase is non-zero.  However, it follows from the
discussion in the previous paragraph that this only happens via terms
proportional to $\sin\theta_{13}$. 

\section{Some Applications}\label{sec:Applications}

The discussed RG effects obviously have important implications
whenever masses and mixings at different energy scales enter
the analysis.

\subsection{Relating the Leptogenesis Parameters to Observations}
\label{sec:Leptogenesis}

One of the most attractive mechanisms for explaining the observed 
baryon asymmetry of the universe, 
$\eta_B = (6.5^{+0.4}_{-0.8}) \cdot10^{-10}$ \cite{Spergel:2003cb},
is leptogenesis \cite{Fukugita:1986hr}.
In this scenario, 
\(\eta_B\) is generated by the out-of-equilibrium 
decay of the same heavy singlet neutrinos which are responsible 
for the suppression of light neutrino masses in the see-saw mechanism.
The masses of the heavy neutrinos are typically assumed to be 
some orders of magnitude below the GUT scale.

Though the parameters entering the leptogenesis mechanism cannot be completely
expressed in terms of low-energy neutrino mass parameters, it is
possible to derive bounds on the neutrino mass scale from the requirement of 
a successful leptogenesis \cite{Buchmuller:2003gz}.
Since, as we demonstrated in Sec.~\ref{sec:RunningOfMasses}, the neutrino 
masses experience corrections of about 20-25\% in the MSSM or 
more than 60\% in the SM, we expect the corrections for such bounds to be 
sizable.

The maximal baryon asymmetry generated in the thermal version of this scenario
is given by \cite{Hamaguchi:2001gw,Davidson:2002qv,Buchmuller:2003gz}
\begin{equation}
 \eta_B^\mathrm{max}
 \,\simeq\,
 0.96\cdot10^{-2}\,\varepsilon_1^\mathrm{max}\,\kappa_\mathrm{f}\;.
\end{equation}
\(\kappa_\mathrm{f}\) is a dilution factor which can be computed
from a set of coupled Boltzmann equations
(see, e.g.\  \cite{Buchmuller:2000as}).
In \cite{Buchmuller:2003gz}, an analytic expression for 
the maximal relevant CP asymmetry was derived,
\begin{equation}
 \varepsilon_1^\mathrm{max} (m_1, m_3, \widetilde{m}_1) 
 \,=\,
 \frac{3}{16\pi}\,\frac{M_1\,m_3}{(v/\sqrt{2})^2}
 \left[1-\frac{m_1}{m_3}
        \left(1+\frac{m_3^2-m_1^2}{\widetilde{m}_1^2}\right)^{1/2}\right]
 ,\label{eq:newEpsilon1max}
\end{equation}
which refines the older bound
\begin{equation}
 \varepsilon_1^\mathrm{max} (m_1, m_3) 
 \,=\,
 \frac{3}{16\pi}\frac{M_1}{(v/\sqrt{2})^2}
 \frac{\Delta m^2_\mathrm{atm}+\Delta m^2_\mathrm{sol}}{m_3}
 \label{eq:oldEpsilon1max}
\end{equation}
and is valid for a normal mass hierarchy in the SM as well as in the 
MSSM.\footnote{
To use these formulae in our conventions for the inverted scheme, one would 
have to replace \((m_1,m_2,m_3)\to(m_3,m_1,m_2)\).}
\(\widetilde{m}_1\) is defined by
\begin{equation}
 \widetilde{m}_1
 =
 \frac{(m_\mathrm{D}^\dagger m_\mathrm{D})_{11}}{M_1}
\end{equation}
with \(m_\mathrm{D}\sim Y_\mathrm{\nu}\) being the neutrino Dirac
mass and typically lies between \(m_1\) and \(m_3\).  It
can be constrained by the requirement of successful leptogenesis
because it controls the dilution of the generated asymmetry.
The authors of \cite{Buchmuller:2003gz} introduced
the `neutrino mass window for baryogenesis' which corresponds 
to the region in the \(\widetilde{m}_1\)-\(M_1\) plane allowing for
successful thermal leptogenesis. The shape and size of the 
`mass window' depends on \(\overline{m}=\sqrt{m_1^2+m_2^2+m_3^2}\), i.e.\ 
it becomes smaller for increasing \(\overline{m}\), and 
\(\overline{m}\ge 0.2\,\mathrm{eV}\) is not compatible with thermal 
leptogenesis.

The calculations relevant for leptogenesis, however, refer to processes at very
high energies, and therefore the RG evolution of the input parameters has to be
taken into account \cite{Barbieri:1999ma}.
The correct procedure would be to assume specific
values for the neutrino mass parameters at low energy, taking into account the
experimental input, evolve them to the scale $M_1$ and test the
leptogenesis mechanism using these values. 
As the full calculation is beyond the scope of this paper, we present
the evolution of the relevant mass parameters, i.e.\  the light neutrino masses,
to the leptogenesis scale \(M_1\) and estimate the size of the error arising 
if RG effects are neglected.  

As discussed in Sec.~\ref{sec:RunningOfMasses}, there are basically
two cases which have to be distinguished, the case of the SM or the
MSSM with small \(\tan\beta\), and the case of the MSSM with large
\(\tan\beta\).

In the first case, running effects can be understood to arise due to
the rescaling of the light neutrino mass eigenvalues under the renormalization 
group. From Eq.~\eqref{eq:oldEpsilon1max} it is clear that the maximal CP asymmetry
scales like the masses. This statement also holds for the asymmetry
from Eq.~\eqref{eq:newEpsilon1max}, if \(\widetilde{m}_1\) is
a linear combination of the light mass eigenvalues.
Hence, the RG yields an enhancement of the CP asymmetry of between 10\% and
80\%, which can be read off from Fig.~\ref{fig:ScalingOfMasses}.
These effects are almost completely independent of the low-energy CP phases.
On the other hand, the dilution factor \(\kappa_\mathrm{f}\) 
is expected to become tiny since larger mass eigenvalues imply
larger Yukawa couplings, which makes the washout more efficient.
This expectation is substantiated by the fact that 
\(\overline{m}\), which controls an important class of
washout processes, also increases under the renormalization group, i.e.\ 
it scales like the masses.
As a detailed numerical calculation of the dilution factor is beyond
the scope of this paper, we refer to \cite{Buchmuller:2000as}, from
which we see that in the region of interest, i.e.\  the edge of the 
mass window, \(\kappa_\mathrm{f}\) decreases exponentially. From 
this behavior, which is also in accordance with the analytic approximations (see, 
e.g.\  \cite{Nielsen:2002pc,DiBari:2002wi}),
we expect that the neutrino mass window
for baryogenesis will rather shrink than become larger when RG effects
are properly taken into account.

\begin{figure}[!htb]
\begin{center}
     \subfigure[\(\tan\beta\)-dependence of the scaling of \(\overline{m}\).\label{fig:RunningMrelTanBeta}]
         {$\CenterEps[1]{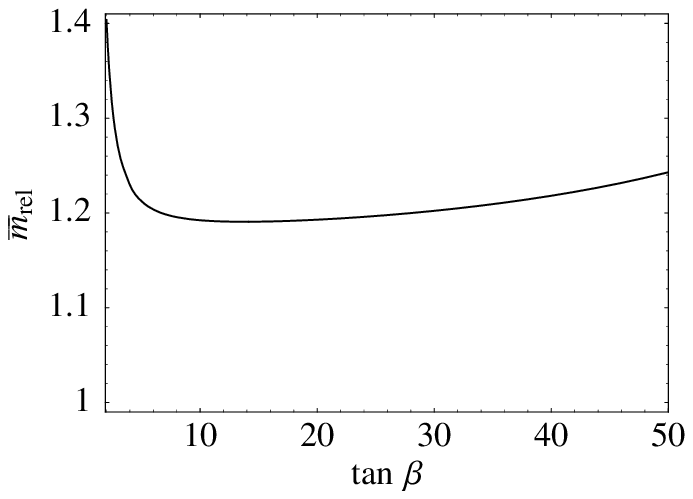}$} 
    \hfil
     \subfigure[$m_1$-dependence of \(\varepsilon_\mathrm{rel}\) for \(\tan\beta=50\).\label{RunningEps}]
                {$\CenterEps[1]{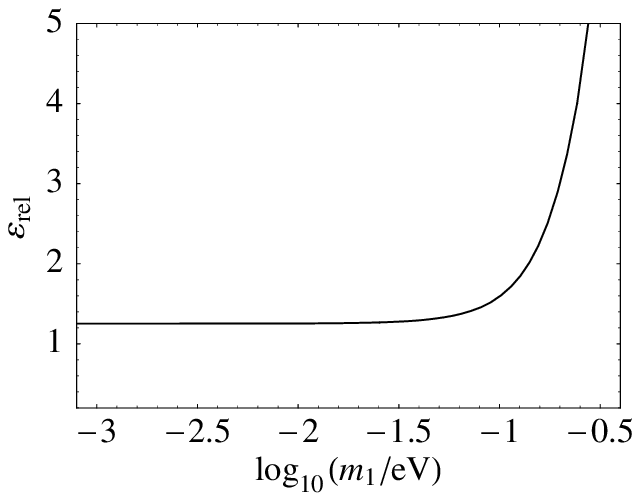}$} 
\end{center}
\vspace*{-5mm}
 \caption{
 Radiative enhancement of \(\overline{m}\) and the CP asymmetry in the MSSM.
 We show only the case of a normal mass hierarchy, since an inverted hierarchy
 yields virtually the same plot.
 We assume a SUSY-breaking scale \(M_\mathrm{SUSY}=1\,\mathrm{TeV}\),
 a leptogenesis scale of \(10^{10}\,\mathrm{GeV}\), and zero phases.
 The mixing angles and mass squared differences are the LMA best-fit values.
 We define
 \(\varepsilon_\mathrm{rel}:=\varepsilon_1^\mathrm{max}(10^{10}\,\mathrm{GeV})/
 \varepsilon_1^\mathrm{max}(M_Z)\).
 In the case of degenerate masses (see the right part of plot (b)),
 \(\varepsilon_1^\mathrm{max}\) can run 
 stronger than the mass eigenvalues since the mass squared differences can have a stronger 
 dependence on the renormalization scale than the squares of the mass eigenvalues
 (cf.\ Fig.~\ref{fig:RunningMasses}).
 } 
 \label{fig:ScalingInMSSM}
\end{figure}

In the second case, i.e.\  in the MSSM for large \(\tan\beta\), 
we distinguish between hierarchical and degenerate mass spectra.
In the hierarchical spectrum, the running of \(\varepsilon_1^\mathrm{max}\) 
is to a high accuracy given by the running of 
\(m_3\),\footnote{For an inverted hierarchy, $m_1$ has to be used instead, whose
 evolution is approximately the same as that of $m_3$ here.}
so that in this case
Fig.~\ref{fig:ScalingOfMasses} yields the relevant plot.
The scaling depends on \(\tan\beta\). In order to illustrate this
dependence, we pick \(M_1=10^{10}\,\mathrm{GeV}\) and plot 
\(\overline{m}_\mathrm{rel}:=\overline{m}(10^{10}\,\mathrm{GeV})/
\overline{m}(M_Z)\) in Fig.~\ref{fig:RunningMrelTanBeta} as a function
of $\tan\beta$, including small values of this parameter as well.  It is clear
that \(\overline{m}\approx m_3\) so that Fig.~\ref{fig:RunningMrelTanBeta}
also shows the scaling of \(\varepsilon_1^\mathrm{max}\). Since 
\(\tan\beta=10\) and \(\tan\beta=50\) correspond to extreme cases,
the scaling factor for different \(M_1\) can be read off from 
Fig.~\ref{fig:ScalingOfMasses} by interpolation.

In the case of a quasi-degenerate mass spectrum (and large \(\tan\beta\)), 
the CP asymmetry can run stronger than the average mass scale because, 
as we already have seen in Sec.~\ref{sec:RunningDm2sol} and 
\ref{sec:RunningDm2atm}, the mass squared differences
can experience a stronger RG enhancement than the squares of the mass 
eigenvalues. We show the evolution of 
\(\varepsilon_\mathrm{rel}:=\varepsilon_1^\mathrm{max}(10^{10}\,\mathrm{GeV})/
\varepsilon_1^\mathrm{max}(M_Z)\) in Fig.~\ref{RunningEps}.
To produce this plot, we employed \eqref{eq:oldEpsilon1max} and inserted
the running mass parameters. For this combination of parameters, 
the low-energy phases do influence the evolution of \(\varepsilon_\mathrm{rel}\) 
by damping its running, and the plot shows the maximal evolution,
which means that the phases are simply set to zero.
The running effects are even larger for the new bound 
\eqref{eq:newEpsilon1max}, since it is more sensitive to
the mass splittings than the old one. More precisely,
for highly degenerate mass spectra it is much smaller than the old one
and the degeneracy can be lifted by running effects.
This strong enhancement of the CP asymmetry may even overcompensate the
decrease of the dilution factor for large $\tan\beta$, so that the
parameter region compatible with thermal leptogenesis grows.

Altogether, we have presented the relevant mass parameters at the scale of
leptogenesis, thus making it convenient to take into account RG 
effects in future studies. Moreover, we have estimated the impact of the
renormalization effects, and found that there are two effects in
opposite directions: The CP asymmetry is enhanced because the mass
squared differences
increase, and the dilution of the baryon asymmetry is more effective
since the overall mass scale rises due to RG effects. As the dependence 
of the dilution factor on the mass scale is stronger than that
of the CP asymmetry, we expect the mass window for baryogenesis
to shrink when RG effects are included in the analysis.
An exception is the case of large $\tan\beta$, where the situation is
more complicated.

Note also that there exist different, non-thermal baryogenesis mechanisms 
\cite{Kumekawa:1994gx} in which the masses of the light neutrinos 
may be almost degenerate \cite{Fujii:2002jw}.
In these kinds of scenarios, RG effects increase the baryon asymmetry,
since \(\varepsilon_1\) increases, while the effects 
from the expected decrease of the dilution factor do not occur.

\subsection{RG Evolution of Bounds on the Neutrino Mass Scale}
The absolute neutrino mass scale at low energy is restricted by
low-energy experiments such as searches for $0\nu\beta\beta$ decay and
cosmological observations.  As usual, the RG evolution of the results
has to be taken into account in order to translate the experimental
results into constraints on high-energy theories.

\subsubsection{Neutrinoless Double Beta Decay}
The amplitude of $0\nu\beta\beta$ decay is proportional to the effective
neutrino mass
\begin{eqnarray}
        \Braket{m_\nu} 
& = &
        (m_\nu)_{11} \ = \ \Bigl| \sum_i U_{1i}^2 \, m_i \Bigr|
\nonumber\\
& = &
        \left|  
         m_1 \, c_{12}^2 c_{13}^2 \, e^{\I \varphi_1} +
         m_2 \, s_{12}^2 c_{13}^2 \, e^{\I \varphi_2} +
         m_3 \, s_{13}^2 \, e^{2\I \delta}
        \right| \;,
\end{eqnarray}
where $U$ is the MNS matrix.
Instead of inserting the lengthy RGEs for all the quantities in the
second line in order to calculate the RG evolution of $\Braket{m_\nu}$,
it is much more convenient to use Eq.~\eqref{eq:BetaKappa}, which 
directly yields
\begin{equation}
        16\pi^2 \: \frac{\D}{\D t}\Braket{m_\nu} 
        \,=\,
        \left( 2 C\, y_e^2 + \alpha \right) \Braket{m_\nu} \;.
\end{equation}
As the first term is negligible, the RG change of the effective neutrino
mass is basically caused by the universal rescaling of the neutrino
masses alone.  It is completely independent of the other neutrino mass
parameters, since neither the running of $y_e$ nor that of the terms in
$\alpha$ is sensitive to them.  Besides, the value of $\tan\beta$ is not
very important here, because $y_e^2$ is always tiny and $\alpha$
contains only the up-type quark Yukawa couplings in the MSSM.
However, there is a dependence on the Higgs mass in the SM.

\begin{figure}[!htb]
 \begin{center}
 \subfigure[SM\label{fig:meffEvolutionSM}]{\CenterEps{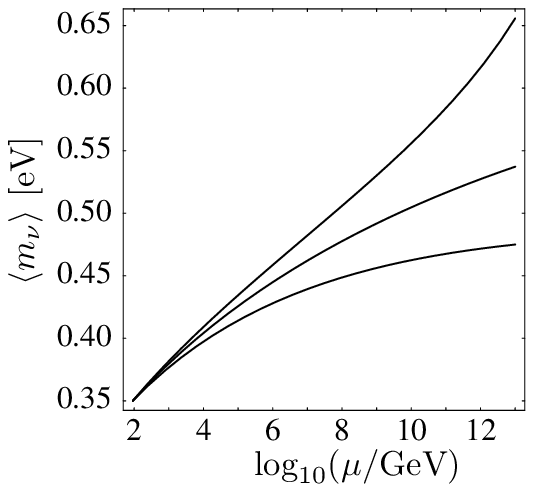}}
 \hfil
 \subfigure[MSSM ($\tan\beta=50$, $M_\mathrm{SUSY}=1.5\,\mathrm{TeV}$)\label{fig:meffEvolutionMSSM}]{\CenterEps{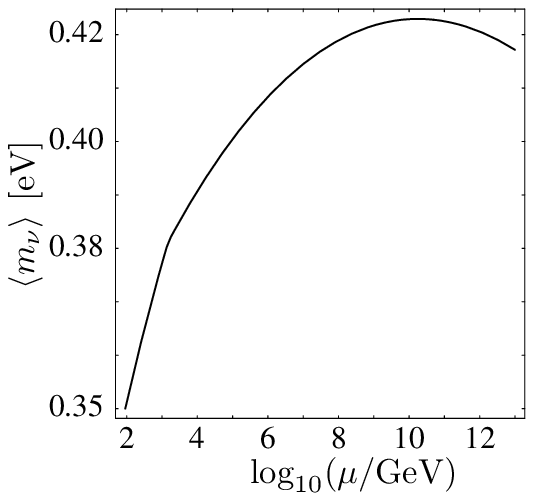}}
 \vspace*{-5mm}
 \end{center}
 \caption{Extrapolation of the experimental upper limit on the effective
  neutrino mass for $0\nu\beta\beta$ decay,
  $\langle m_\nu \rangle = 0.35\,\mathrm{eV}$, to higher energies.  
  The SM curves correspond to Higgs masses of
  \(114\,\mathrm{GeV}\), \(165\,\mathrm{GeV}\) and 
  \(190\,\mathrm{GeV}\) (from bottom to top).  In the MSSM, a
  light Higgs mass of \(120\,\mathrm{GeV}\) is used.
 }
 \label{fig:EffectiveMassEvolution1}
\end{figure}

Currently, the best experimental upper limit on the effective neutrino
mass is about $\Braket{m_\nu}<0.35\,\mathrm{eV}$
\cite{Klapdor-Kleingrothaus:2000sn,Aalseth:2002rf},
with some uncertainty due to nuclear matrix elements.
Fig.~\ref{fig:EffectiveMassEvolution1} shows the running of this limit
in the SM and the MSSM.  As it is very close to the best-fit value of
the recently claimed evidence for double beta decay,
$\Braket{m_\nu}=0.39\,\mathrm{eV}$ \cite{Klapdor-Kleingrothaus:2001ke}, the
evolution of the latter is nearly identical.  The SM plot contains
three curves corresponding to different Higgs masses in the current
experimentally allowed region.  In the MSSM, the light Higgs mass is
chosen to be about $120\,\mathrm{GeV}$.
The running is much more significant in the SM than in the MSSM because of
the contribution of the Higgs self-coupling.

\subsubsection{WMAP Bound}
\begin{figure}[!htb]
 \begin{center}
 \subfigure[SM\label{fig:WMAPEvolutionSM}]{\CenterEps{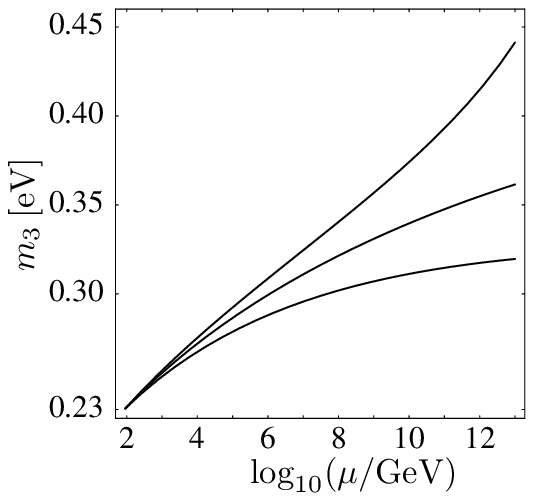}}
 \hfil
 \subfigure[MSSM ($\tan\beta=50$, $M_\mathrm{SUSY}=1.5\mathrm{TeV}$)\label{fig:WMAPEvolutionMSSM}]{\CenterEps{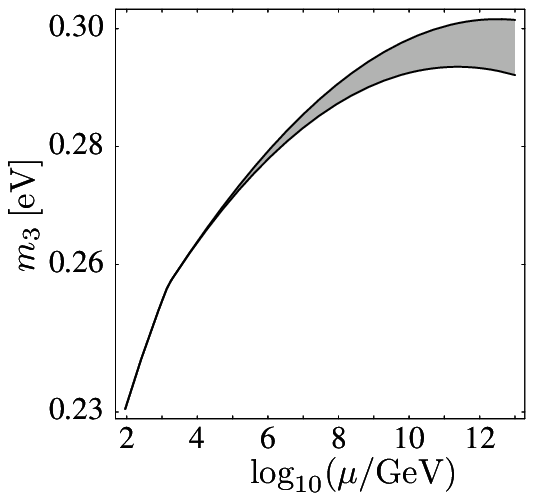}}
 \vspace*{-5mm}
 \end{center}
 \caption{
  Extrapolation of the upper limit on the neutrino mass from WMAP,
  $m_i \lesssim 0.23\,\mathrm{eV}$ to higher energies, 
  represented by the running of the mass eigenvalue $m_3$.  
  The SM curves correspond to Higgs masses of
  \(114\,\mathrm{GeV}\), \(165\,\mathrm{GeV}\) and 
  \(190\,\mathrm{GeV}\) (from bottom to top).  In the MSSM, a
  light Higgs mass of \(120\,\mathrm{GeV}\) is used.
 }
 \label{fig:WMAPEvolution}
\end{figure}

Combining the observations of the cosmic microwave background by the
WMAP satellite with other astronomical data allows to place an upper
bound of about $0.7\,\mathrm{eV}$ onto the sum of the light neutrino masses
\cite{Spergel:2003cb}. This implies
\begin{equation}
        m_i \lesssim 0.23\, \text{eV}
\end{equation}
for each mass eigenvalue.  Analogous to the limit from $0\nu\beta\beta$ 
decay in the previous section, this bound is modified substantially by
the RG evolution.  This is shown in Fig.~\ref{fig:WMAPEvolution} for
the eigenvalue $m_3$. As discussed in Sec.~\ref{sec:RunningOfMasses},
the running of the mass eigenvalues is not sensitive to the mixing
parameters in the SM, but it depends on the Higgs mass.  In the MSSM,
the variation of the phases causes a slight modification of the running,
but its order of magnitude is only a few percent even for the large
$\tan\beta$ used in the plot.  The influence of $\theta_{13}$ is
negligible.  Interestingly, the evolution of the sum of the mass
eigenvalues is virtually independent of the mixing parameters for nearly
degenerate neutrinos both in the SM and in the MSSM.  This can be
explained by considering the sum of the RGEs
\eqref{eq:EvolutionOfMassEigenvalues}.  For $m_1 \sim m_2 \sim m_3$, the
terms proportional to $y_\tau^2$ add up to 1, with small corrections of
the order of $\frac{\Delta m^2_\mathrm{atm}}{m^2}$ and $\theta_{13}$.

\subsection{Constraints on Neutrino Properties from RG Effects}
\label{sec:ConstraintsFromRG}

One may wonder if deviations from $\theta_{13}=0$ and $\theta_{23}=\pi/4$ 
exist which are the consequence of radiative corrections. Let us assume therefore 
that \(\theta_{13}=0\) or $\theta_{23}=\pi/4$ are given by some high-energy 
model. Low-energy deviations from the exact values are then RG effects, 
which can be compared to the sensitivities of future experiments. Therefore 
we investigate in a model-independent way the size of RG corrections 
to $\theta_{13}$ and $\theta_{23}$ from the running of the effective neutrino 
mass operator between the see-saw scale and the electroweak scale. 

\subsubsection{Corrections to $\boldsymbol{\theta_{13}}$}

As pointed out in Sec.~\ref{sec:RunningTheta13}, it is a rather good approximation
to assume \(\Dot{\theta}_{13}\simeq\)~const.\ in Eq.~\eqref{eq:Theta13Dot}, 
which leads to an RG evolution with a constant slope depending on 
the Dirac CP phase $\delta$ and the Majorana phases $\varphi_1$ and $\varphi_2$.
Therefore, let us first apply the naive estimate \eqref{eq:NaiveRGChange} 
explicitly to the change of $\theta_{13}$ in the MSSM for nearly
degenerate neutrinos.  In this case, the enhancement factor
$m^2/\Delta m^2_\mathrm{atm}$ leads to a generic change of
$\theta_{13}$ under the RG that exceeds the detection limit of future
experiments even for moderate values of $\tan\beta$.  For example,
$m_1=0.1\,\mathrm{eV}$ and $\tan\beta=30$ yield a change in 
$\sin^2 2\theta_{13}$ of $\Delta\!\sin^2 2\theta_{13} \sim 0.5\cdot10^{-2}$,
which is further enhanced by a factor of 4 if the Majorana phases are
aligned properly.

\begin{figure}[p]
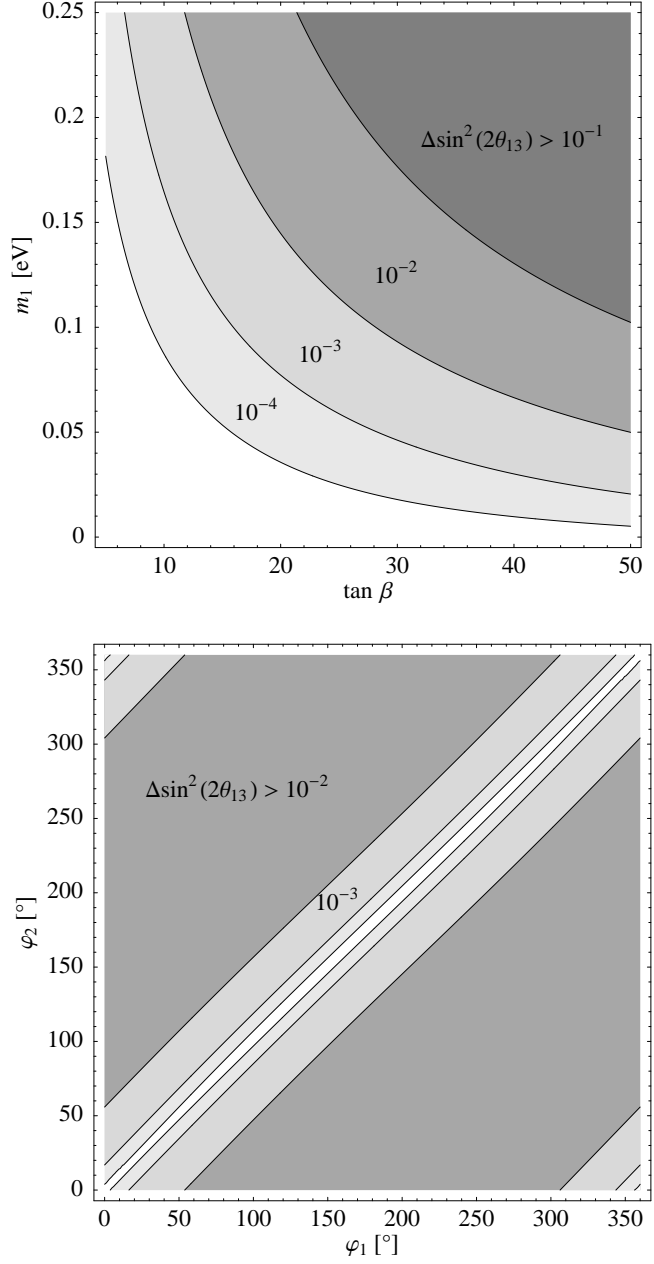

        \begin{center}
        {$\CenterEps[0.75]{RGCorrectionTheta13}$}\\ \vspace{0.5cm}\hspace{0.2cm}
        {$\CenterEps[0.75]{RGCorrectionTheta13_Phases}$}
 \vspace*{-3mm}
 \end{center}
 \caption{\label{fig:RGCorrt13}\label{fig:RGCorrt13_Phases}%
 Corrections to $\theta_{13}$ from the RG evolution 
 between $10^2$ and $10^{12}\,\mathrm{GeV}$ in the MSSM, calculated
 using the analytical approximations with initial conditions
 $\theta_{13}=0$ and LMA best-fit values for the remaining parameters.  
 The upper diagram shows the dependence on 
 $\tan \beta$ and on the mass of the lightest neutrino
 for the case of a normal mass hierarchy and phases 
 $\varphi_1=0$ and $\varphi_2=\pi$.
 In the lower diagram the dependence on the Majorana phases $\varphi_1$ and 
 $\varphi_2$ is shown for $\tan\beta=50$ and $m_1=0.08\,\mathrm{eV}$. 
 The contour lines are defined as in the upper diagram. 
 In order to apply Eq.~\eqref{eq:AnalyticApproxT13} to
 the case $\theta_{13}=0$, where $\delta$ is undefined, 
 the analytic continuation of Eq.~\eqref{eq:DeltaZeroT13} has been used. 
 }
\end{figure}

In order to obtain a more detailed picture, we now apply
Eq.~\eqref{eq:AnalyticApproxT13} to calculate the RG correction to the
initial value $\theta_{13}=0$ between some high energy
scale $M_1$, where neutrino masses are generated, and low energy, 
i.e.\ $10^2\,\mathrm{GeV}$. 
In this case the initial value of the Dirac phase $\delta$ is determined
by the analytic continuation Eq.~\eqref{eq:DeltaZeroT13}. 
For the examples we take $M_1 = 10^{12}\,\mathrm{GeV}$.
The approximate size of the RG corrections to $\sin^2 2\theta_{13}$
in the MSSM is shown in Fig.~\ref{fig:RGCorrt13}. In the upper diagram 
it is plotted as a
function of $\tan\beta$ and the lightest neutrino mass $m_1$ for constant
Majorana phases $\varphi_1=0$ and $\varphi_2=\pi$.  The lower diagram
shows the dependence of the corrections on $\varphi_1$ and $\varphi_2$
for $\tan\beta=50$ and $m_1=0.08\,\mathrm{eV}$ in the case of a normal
mass hierarchy. The diagrams look rather similar for an inverted
hierarchy.
Analytically, the pattern of the upper plot is easy to understand, and for the
lower one there is a simple explanation as well.  Consider partially or
nearly degenerate neutrino masses.  Then Eq.~\eqref{eq:Theta13Dot} yields
to a reasonably good approximation
\begin{eqnarray} \label{eq:T13DotApprox} 
\Dot{\theta}_{13}
& \approx & 
        \frac{C y_\tau^2}{32\pi^2} \, 
        \sin 2\theta_{12} \, \sin 2\theta_{23} \,
        \frac{m^2}{\Delta m^2_\mathrm{atm}}
        \left[
         \cos(\varphi_1-\delta) - \cos(\varphi_2-\delta)
        \right]
\nonumber\\*
& \propto &
        \sin\frac{\varphi_1+\varphi_2-2\delta}{2} \,
        \sin\frac{\varphi_1-\varphi_2}{2} \;.
\end{eqnarray}
Applying an analogous approximation to Eq.~\eqref{eq:DeltaZeroT13}, it
can easily be shown that the first term in the second line is always 
$\pm 1$, so that the running is completely determined
by the difference of the Majorana phases.  This leads to the diagonal
bands in Fig.~\ref{fig:RGCorrt13_Phases}, in particular the white one
corresponding to $\varphi_1-\varphi_2=0$.  If one starts with a small
but non-zero $\theta_{13}$, which allows an arbitrary $\delta$, it turns
out that the RG evolution quickly drives $\delta$ to a value
satisfying Eq.~\eqref{eq:DeltaZeroT13}, so that the final pattern of
Fig.~\ref{fig:RGCorrt13_Phases} is unchanged.

Planned reactor experiments \cite{Huber:2003pm} and next generation 
superbeam experiments \cite{Huber:2002rs,Minakata:2003ca} are expected 
to have an approximate sensitivity on $\sin^2 2\theta_{13}$ of $10^{-2}$. 
From Fig.~\ref{fig:RGCorrt13} we find that the radiative corrections exceed 
this value for large regions of the currently allowed parameter space, 
unless there are cancellations due to Majorana phases, i.e.\ 
$\varphi_1\approx\varphi_2$ (which might be due to some symmetry).
If so, the effects are generically smaller than $10^{-2}$ as can be seen 
from the lower diagram. Future upgraded superbeam experiments like 
JHF-HyperKamiokande have the potential to further push
the sensitivity to about $10^{-3}$ and with a neutrino factory
even about $10^{-4}$ might be reached. 

From the theoretical point of view, one would expect that even if some 
model predicted $\theta_{13}=0$ at the energy scale of neutrino mass  
generation, RG effects would at least produce a non-zero
value of the order shown in Fig.~\ref{fig:RGCorrt13}. 
Consequently, experiments with such a sensitivity have a large 
discovery potential for $\theta_{13}$.
We should point out that this is a conservative estimate, since 
if neutrino masses are e.g.\ determined by GUT scale physics,
model-dependent radiative corrections in the region between $M_1$ and 
$M_\mathrm{GUT}$ contribute as well 
\cite{Casas:1999tp,Casas:1999ac,King:2000hk,Antusch:2002rr,Antusch:2002hy,Antusch:2002fr} and there       
can be additional corrections from physics above the GUT scale 
\cite{Vissani:2003aj}.
On the other hand, if experiments do not measure $\theta_{13}$, this will 
improve the upper bound on $\theta_{13}$. Parameter space regions where the
corrections are larger than this bound will then appear unnatural from 
the theoretical side.

\subsubsection{Corrections to $\boldsymbol{\theta_{23}}$}
We now consider the RG corrections which induce a deviation of $\theta_{23}$
from $\pi/4$, even if some model predicted this specific value at high
energy.  We apply the analytical formula \eqref{eq:Theta23Dot} with a
constant right-hand side in order to calculate the running in the MSSM 
between $M_Z$ and the see-saw scale, which we take as 
$M_1 = 10^{12}\,\mathrm{GeV}$ for our examples.
As initial conditions we assume small $\theta_{13}$ at $M_1$ and low-energy 
best-fit values for the remaining lepton mixings and the neutrino mass 
squared differences. 
In leading order in $\theta_{13}$, the evolution is of course 
independent of the Dirac phase $\delta$. 

\begin{figure}[p]
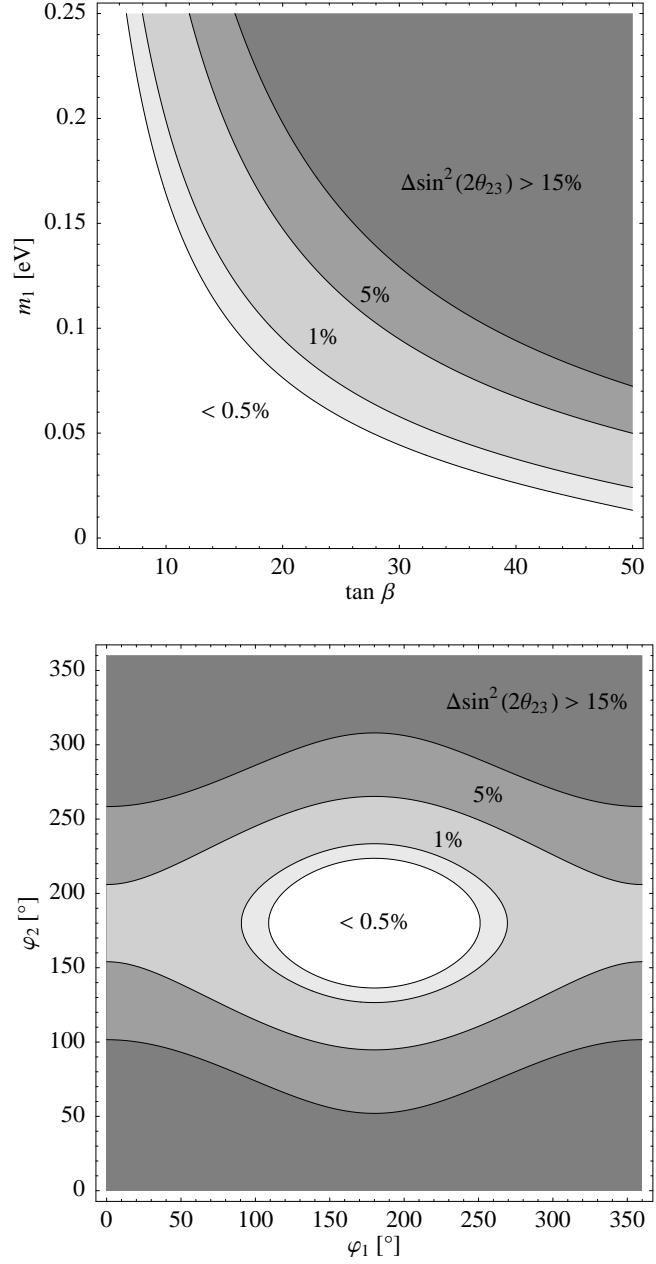

        \begin{center}
        {$\CenterEps[0.75]{RGCorrectionTheta23}$}\\ \vspace{0.5cm}\hspace{0.2cm}
        {$\CenterEps[0.75]{RGCorrectionTheta23_Phases}$}
 \vspace*{-3mm}
 \end{center}
 \caption{\label{fig:RGCorrt23}\label{fig:RGCorrt23_Phases}%
 Corrections to $\theta_{23}$ from the RG evolution 
 between $10^2\,\mathrm{GeV}$ and $10^{12}\,\mathrm{GeV}$ in the MSSM,
 calculated from the analytical approximation Eq.~\eqref{eq:Theta23Dot} 
 with initial conditions
 $\theta_{23}=\pi/4$, small $\theta_{13}=0$ and LMA best-fit values for the remaining
 parameters. 
 The upper diagram shows the dependence on 
 $\tan \beta$ and on the mass $m_1$ of the lightest neutrino
 for the case of a normal mass hierarchy and phases 
 $\varphi_1=\varphi_2=0$.
 In the lower diagram the dependence on the Majorana phases $\varphi_1$ and 
 $\varphi_1$ is shown for the example $\tan \beta =50$ and $m_1=0.1$. 
 Note that for small $\theta_{13}$ the results are
 independent of the Dirac phase to a good approximation.
 }
\end{figure}

The size of the RG corrections in the MSSM is shown in
Fig.~\ref{fig:RGCorrt23}.  From the upper diagram it  
can be read off for desired values of $\tan\beta$ and the lightest mass
eigenvalue $m_1$ in an example with vanishing Majorana phases.
The lower diagram shows its dependence on the Majorana phases $\varphi_1$ and 
$\varphi_2$ for $\tan\beta=50$, $m_1=0.1\,\mathrm{eV}$ and a normal mass
hierarchy. The diagrams look rather similar in the case of an inverted hierarchy. 
The effects of the Majorana phases can easily be understood from 
Eq.~\eqref{eq:Theta23Dot}.
In the region with $\varphi_1 \approx \varphi_2 \approx \pi$
(again, this might be, e.g., due to some symmetry), both $|m_2\, e^{\I \varphi_2} + m_3|^2$ 
and $|m_1\, e^{\I \varphi_1} + m_3|^2$  are small for quasi-degenerate 
neutrinos, which gives the ellipse with small radiative corrections in the
center of the lower diagram. 
Such cancellations are not possible with hierarchical masses, but the RG
effects are generally not very large in this case, as shown by the upper
plot.

Even if a model predicted $\theta_{23}=\pi/4$ at some high energy
scale, we would thus expect
radiative corrections to produce at least a deviation from this
value of the size shown in Fig.~\ref{fig:RGCorrt23}, so that experiments
with such a sensitivity are expected to measure a deviation of 
$\theta_{23}$ from $\pi/4$.
The sensitivity to $\sin^2 2\theta_{23}$ of future superbeam 
experiments like JHF-SuperKamiokande is expected to be approximately 
1\% (see e.g.\ \cite{Itow:2001ee}). This can now be compared with 
Fig.~\ref{fig:RGCorrt23}. We find that the radiative corrections exceed this
value for large regions of the currently allowed parameter space, where
no significant cancellations due to Majorana phases occur.  This means
that $\varphi_1$ and $\varphi_2$ must not be too close to $\pi$.
Otherwise, the effects are generically smaller as can be seen from 
the lower diagram. Upgraded superbeam experiments or a neutrino factory 
might even reach a sensitivity of about $0.5$\%. 
As argued for the case of $\theta_{13}$, if experiments measure
$\theta_{23}$ rather close to $\pi/4$, parameter combinations implying
larger radiative corrections than the measured deviation will appear
unnatural from the theoretical point of view. 

\clearpage

\section{Conclusions} 
\label{sec:Conclusions}

We have derived compact expressions which allow an analytical understanding
of the running of neutrino masses, leptonic mixing angles and CP phases
in the SM and MSSM. The results are given directly in terms of these quantities
as well as gauge and Yukawa couplings, and especially for a small angle 
$\theta_{13}$ the expressions become very simple, even when non-vanishing CP phases are present. We have extensively 
compared those formulae to numerical results and we have found that the RG
evolution of the physical parameters is described qualitatively, and to a 
reasonable accuracy also quantitatively, very well.
We have shown that Dirac and Majorana CP phases can have a drastic influence 
on the RG evolution of the mixing parameters. 
We have reproduced and illustrated some effects that were previously
described in the literature.  As a particularly interesting example, we
have discussed the radiative generation of the Dirac phase from the
Majorana phases.  Besides, we have derived new results,
for example concerning the running of the CP phases.
Even though the RG effects for the mixing
parameters in the SM are rather small, the RG effects for the masses are not, 
and have to be taken into account in any careful analysis which relates high 
and low energy scales. In the MSSM, especially for large $\tan\beta$, the 
evolution of the mixings and phases can be large.

The RG evolution has interesting phenomenological implications. In the case of
leptogenesis, we have estimated the corrections which arise if the running 
is appropriately taken into account and found that the mass window for 
baryogenesis is likely to shrink when those corrections are considered.
In order to simplify the inclusion of RG effects in future calculations, 
we provide the relevant information of the mass parameters at the
leptogenesis scale. Furthermore, we investigated the extrapolation of the 
upper bounds on the neutrino mass scale from \(0\nu\beta\beta\) decay
experiments and WMAP to higher energy scales, where they become restrictions 
for model building. Experimentally one finds \(\theta_{23}\simeq \pi/4\), 
\(\theta_{13}\simeq 0\). The deviations from $\pi/4$ and zero may have a 
radiative origin and we calculated therefore in a model-independent analysis the
RG corrections to \(\theta_{23}=\pi/4\), \(\theta_{13}=0\). With future 
precision experiments this may lead to interesting insights into model 
parameters.

To conclude, we have obtained analytic formulae which are a useful tool to understand 
the RG corrections, relevant whenever parameters at two different energy
scales are compared. This has been demonstrated in the phenomenological applications.

\section*{Acknowledgments}

This work was supported in part by the 
``Sonderforschungsbereich~375 f\"ur Astro-{}Teil\-chen\-phy\-sik der 
Deutschen Forschungsgemeinschaft''.
We would like to thank W. Buch\-m\"ul\-ler, K.~Hamaguchi, P.~Huber,
R.~N.~Mohapatra and W.~Winter for interesting discussions.

\section*{Appendix}
\appendix

\renewcommand{\thesection}{\Alph{section}}
\renewcommand{\thesubsection}{\Alph{section}.\arabic{subsection}}
\def\theequation{\Alph{section}.\arabic{equation}}
\renewcommand{\thetable}{\arabic{table}}
\renewcommand{\thefigure}{\arabic{figure}}
\setcounter{section}{0}
\setcounter{equation}{0}

\section{Definition and Extraction of Mixing Parameters}
\label{app:MixingParameters}

\subsection{Standard Parametrization}
In this section we describe our conventions and how mixing angles and 
phases can be extracted from mass matrices.
For a general unitary matrix we choose the so-called 
standard-parametrization
\begin{eqnarray}\label{eq:StandardParametrizationU}
 U & = &\diag(e^{\I\delta_{e}},e^{\I\delta_{\mu}},e^{\I\delta_{\tau}}) \cdot V \cdot 
 \diag(e^{-\I\varphi_1/2},e^{-\I\varphi_2/2},1)
\end{eqnarray}
where 
\begin{equation}
 V=\left(
 \begin{array}{ccc}
 c_{12}c_{13} & s_{12}c_{13} & s_{13}e^{-\I\delta}\\
 -c_{23}s_{12}-s_{23}s_{13}c_{12}e^{\I\delta} &
 c_{23}c_{12}-s_{23}s_{13}s_{12}e^{\I\delta} & s_{23}c_{13}\\
 s_{23}s_{12}-c_{23}s_{13}c_{12}e^{\I\delta} &
 -s_{23}c_{12}-c_{23}s_{13}s_{12}e^{\I\delta} & c_{23}c_{13}
 \end{array}
 \right)
\end{equation}
with \(c_{ij}\) and \(s_{ij}\) defined as \(\cos\theta_{ij}\) and
\(\sin\theta_{ij}\), respectively. 

\subsection{Extracting Mixing Angles and Phases}
\label{sec:ExtractingMixingAngles}

In this standard-parametrization, the mixing angles \(\theta_{13}\) 
and \(\theta_{23}\) can be chosen to lie between \(0\) and \(\frac{\pi}{2}\),
and by reordering the masses, $\theta_{12}$ can be restricted to \(0\le\theta_{12}\le\frac{\pi}{4}\).
For the phases the range between \(0\) and \(2\pi\) is required.
In order to read off the mixing parameters, we use the following procedure:
\begin{enumerate}
 \item \(\theta_{13}=\arcsin(|U_{13}|)\).
 \item \(\displaystyle \theta_{12}=\left\{\begin{array}{ll}
 \displaystyle \arctan\left(\frac{|U_{12}|}{|U_{11}|}\right) \quad
        & \text{if}\;U_{11}\ne0\\
 \frac{\pi}{2} & \text{else}
 \end{array}\right.\)
 \item \(\displaystyle \theta_{23}=\left\{\begin{array}{ll}
 \displaystyle \arctan\left(\frac{|U_{23}|}{|U_{33}|}\right) \quad
        & \text{if}\;U_{33}\ne0\\
 \frac{\pi}{2} & \text{else}
 \end{array}\right.\)
 \item \(\delta_\mu = \arg(U_{23})\)
 \item \(\delta_\tau = \arg(U_{33})\)
 \item \label{step6}\(\displaystyle\delta=
 -\arg\left(\frac{\displaystyle\frac{U_{ii}^*U_{ij}U_{ji}U_{jj}^*}
        {c_{12}\,c_{13}^2\,c_{23}\,s_{13}}
        +c_{12}\,c_{23}\,s_{13}}
        {s_{12}\,s_{23}}\right)\)\\
 where \(i,j\in\{1,2,3\}\) and \(i\ne j\).
 \item \(\delta_e=\arg(e^{\I\delta}\,U_{13})\)
 \item \(\displaystyle\varphi_1=2\arg(e^{\I\delta_e}\,U_{11}^*)\)
 \item \label{step9}\(\displaystyle\varphi_2=2\arg(e^{\I\delta_e}\,U_{12}^*)\)
\end{enumerate}
Here we used the relation
\begin{eqnarray}
 U_{ii}^*U_{ij}U_{ji}U_{jj}^*
 & = &
 c_{12}\,c_{13}^2\, 
 c_{23}\,s_{13}
 \left(e^{-\I\delta}\,s_{12}\,s_{23} - c_{12}\,c_{23}\,s_{13}\right)
 \;,\nonumber
\end{eqnarray}
which holds for \(i,j\in\{1,2,3\}\) and \(i\ne j\).
Note that this relation is often used in order to introduce 
the Jarlskog invariants \cite{Jarlskog:1985ht}
\begin{eqnarray}
 J_\mathrm{CP} 
 & = &
 \frac{1}{2} \left| \im (U_{11}^*U_{12}U_{21}U_{22}^*)\right|
 \,=\, 
 \frac{1}{2} \left| \im (U_{11}^*U_{13}U_{31}U_{33}^*)\right|
 \nonumber\\
 & =  &
 \frac{1}{2} \left| \im (U_{22}^*U_{23}U_{32}U_{33}^*)\right|
 \, = \,
 \frac{1}{2}\left|c_{12}\,c_{13}^2\,
    c_{23}\,\sin \delta \,
    s_{12}\,s_{13}\,
    s_{23}\right|\;.
\end{eqnarray}
For the sake of a better numerical stability, one can choose any of the three
combinations. In particular, if the modulus of one of the \(U_{ij}\) is very
small, it turns out to be more accurate to choose a combination in which this
specific \(U_{ij}\) does not appear.

\subsection{Leptonic Mixing Matrix} \label{sec:LeptMixingMatrix}

\begin{figure}[h]
 \begin{center}
  \begin{subfigure}[Normal mass hierarchy.]{\label{subfig:NeutrinoMassScheme1}
        $\begin{array}{c}\CenterEps{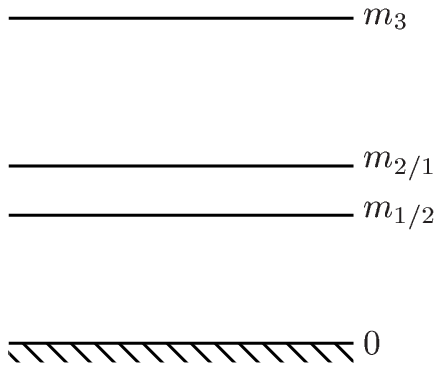}\\[1cm]\phantom{\text{M\"ummel}}
        \end{array}$
   }
  \end{subfigure}
  \hfil
  \begin{subfigure}[Inverted mass hierarchy.]{\label{subfig:NeutrinoMassScheme2}
  $\begin{array}{c}\CenterEps{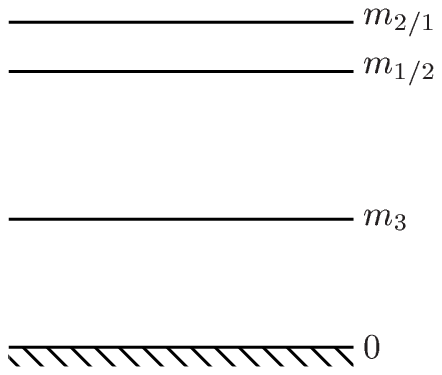}\\[1cm]\phantom{\text{M\"ummel}}
  \end{array}$
   }
  \end{subfigure}
 \end{center}
 \vspace*{-5mm}
 \caption{The normal and inverted mass hierarchy.}
 \label{fig:NeutrinoMassSchemes}
 \vspace*{5mm}
\end{figure}

Since the effective neutrino mass matrix is symmetric, 
it can be diagonalized by a unitary matrix \(U_\nu\),
\begin{equation}
  U_\nu^T\, m_\nu\, U_\nu\,=\, \diag(m_1,m_2,m_3) \;.
\end{equation}
The form of \(U\) depends on a prescription how to order the mass eigenvalues.
In order to obtain a mixing matrix which can be compared with the experimental data, 
the choice of the prescription is somewhat subtle. From experiment we know that there is a
small mass difference, called \(\Delta m^2_\mathrm{sol}=m_i^2-m_j^2\), and a larger one, referred to as 
\(\Delta m^2_\mathrm{atm}=m_k^2 - m_\ell^2\). By convention, the masses are labeled such
that \(i,j\ne 3\) while either \(k\) or \(\ell\) equals 3. 
The different schemes are depicted in Fig.~\ref{fig:NeutrinoMassSchemes}.
The mass label 2 is attached to the
eigenvector with the lower modulus of the first component. We are doing this
since we want to read off a mixing angle \(\theta_{12}\) less then \(45^\circ\).

The neutrino mixing matrix \(U_\mathrm{MNS}\) can then be read off in the following way:
\begin{enumerate}
 \item Diagonalize \(Y_e^\dagger Y_e\) by \(U_e\), i.e.\  
  \(Y_e\to U_e^\dagger \cdot Y_e^\dagger\cdot Y_e\cdot U_e
  =\diag\left(y_e^2,y_\mu^2,y_\tau^2\right)\) where 
  \(y_f^2\) are positive for \(f\in\{e,\mu,\tau\}\).
 \item Change the basis according to
  \(m_\nu\to m_\nu'=U_e^T \cdot m_\nu \cdot U_e\). 
 \item Diagonalize \(m_\nu'\): 
  \(m_\nu'\to U_\mathrm{MNS}^T\cdot m_\nu'\cdot U_\mathrm{MNS}=\diag(m_1,m_2,m_3)\)
  where \(m_i>0\).
\end{enumerate}
Then \(U_\mathrm{MNS}\) contains the leptonic mixing angles
which can be read off as described in Sec.~\ref{sec:ExtractingMixingAngles}. 
Note that \(m_1<m_2<m_3\) is not necessarily fulfilled, as we already mentioned before
(cf.\ Fig.~\ref{fig:NeutrinoMassSchemes}).

\section{Derivation of the Analytical Formulae}
\label{sec:DerivRGEs}

To derive the RGEs for the mixing parameters, we follow in general the methods of
\cite{Babu:1987im}. The RGE for \(\kappa\) reads
\begin{equation}\label{eq:GeneralRGEforKappa}
 16\pi^2\,\frac{\D\kappa}{\D t}
 \,=\,
 \alpha\,\kappa + P^T\,\kappa + \kappa\, P\;,
\end{equation}
where all terms with trivial flavour structure are absorbed in \(\alpha\).
\(\kappa\) can be diagonalized (in the basis where \(Y_e\) is diagonal)
by a unitary transformation,
\begin{equation}
 U(t)^T\,\kappa(t)\, U(t) 
 \,=\, D(t) 
 \,=\, \frac{4}{v^2} \,\diag\big(m_1(t),m_2(t),m_3(t)\big)\;.
\end{equation}
We hence obtain
\begin{eqnarray}
 \lefteqn{
        \frac{\D}{\D t}\left(U^*\, D\, U^\dagger\right)
        =\Dot U^*\, D \, U^\dagger + U^*\, D \, \Dot U^\dagger
        +U^*\, \Dot D\, U^\dagger
        }
 \nonumber\\
 & \stackrel{\eqref{eq:GeneralRGEforKappa}}{=} &        
 \frac{1}{16\pi^2}\,
 \left(\alpha\,U^*\, D\, U^\dagger
        +P^T\, U^*\, D \, U^\dagger
        +U^*\, D\, U^\dagger \, P\right)
 \;.
\end{eqnarray}
Multiplying with \(U^T\) from the left and with \(U\) from the right yields
\begin{equation}
 U^T\, \Dot U^* \, D
 +D\, \Dot U^\dagger \, U
 + \Dot D
 \,=\,
 \frac{1}{16\pi^2}
 \left[\alpha\,D+P^{\prime\,T}\, D + D\, P'\right]
 \;,
\end{equation}
where we have introduced \(P'=U^\dagger\, P\, U\).
The next step is defining an anti-Hermitian matrix \(T\) by
\begin{equation}\label{eq:EvolutionOfU}
 \frac{\D}{\D t} U \,=\, U\, T\;.
\end{equation}
With this definition, we find
\begin{equation}\label{eq:MasterEquationForDTP}
 \Dot D
 \,=\,
 \frac{1}{16\pi^2}\,
 \left(\alpha\,D + P^{\prime\,T}\, D + D\, P'\right)
 -T^*\, D+ D\, T
 \;,
\end{equation}
where the anti-hermiticity of \(T\) was used.
Since the left-hand side of this equation is diagonal and real per definition,
the right-hand side has to possess these properties as well,
\begin{equation}\label{eq:DotMi1}
 \Dot m_i
 \,=\,
 \frac{1}{16\pi^2}\,
 \left(\alpha\, m_i+ 2\,P'_{ii}\,m_i\right)
 +(T_{ii}-T_{ii}^*)\,m_i
 \;.
\end{equation}
Note that here and in the following equations, no sum over repeated
indices is implied.
The second bracket is purely imaginary, hence it has to cancel with the
imaginary part of the first one,
\begin{equation}\label{eq:ImDiagT}
 2\,\im T_{ii}
 \,=\,
 \frac{-1}{16\pi^2}\,(\im\alpha +2\im P'_{ii})
 \;,
\end{equation}
and we further confirm eq.\ (15) of \cite{Casas:1999tg},
which translates with our conventions to
\begin{equation}\label{eq:EvolutionOfEffectiveMasses}
 16\pi^2\,\Dot m_i
 \,=\,
 \left(\re \alpha+2\,\re P'_{ii}\right)\,m_i
 \;.
\end{equation}
Eq.~\eqref{eq:ImDiagT} differs from Eq.~(19) of \cite{Casas:1999tg}, 
where the imaginary part of \(\alpha\) is not present;
however, this difference is irrelevant in the SM and the MSSM, where 
$\alpha$ is real.
By comparing the off-diagonal parts of \eqref{eq:MasterEquationForDTP}
we find
\begin{equation} \label{eq:TPprimeIntermediate}
 m_i\,T_{ij}-T_{ij}^*\,m_j
 \,=\,
 -\frac{1}{16\pi^2}\,
 \left(P^{\prime\,T}_{ij}\,m_j+m_i\,P'_{ij}\right)
 \;.
\end{equation}
Adding and subtracting this equation and its complex conjugate,
we obtain for \(i\ne j\)
\begin{subequations}\label{eq:ExpressTbyPprime}
\begin{eqnarray}
 16\pi^2\,\re T_{ij}
 & = &
 -\frac{m_j\,\re P'_{ji}+m_i\,\re P'_{ij}}{m_i-m_j}
 \;,
 \\
 16\pi^2\,\im T_{ij}
 & = &
 -\frac{m_j\,\im P'_{ji}+m_i\,\im P'_{ij}}{m_i+m_j}
 \;.
 \label{eq:ImTijBelow}
\end{eqnarray}
\end{subequations}
Let us now focus on Hermitian \(P\), which implies Hermitian
\(P'\), for a moment. Using \(\re P'_{ji}=\re P^{\prime\,*}_{ij}
=\re P'_{ij}\) and an analogous relation for \(\im P'_{ij}\),
we obtain in this case
\begin{subequations}
\begin{eqnarray}
 16\pi^2\,\im T_{ij}
 & = &
 -\frac{m_i-m_j}{m_i+m_j}\,\im P'_{ij}
 \;,
 \\
 16\pi^2\,\re T_{ij}
 & = &
 -\frac{m_i+m_j}{m_i-m_j}\,\re P'_{ij}
 \;.
\end{eqnarray}
\end{subequations}
In order to obtain the renormalization group equations for the mixing angles,
we use \eqref{eq:EvolutionOfU},
\begin{equation}\label{eq:EvolutionOfU2}
 U^\dagger\, \Dot U = T\;.
\end{equation}
Inserting the standard parametrization 
\eqref{eq:StandardParametrizationU}, we can express the left-hand side
of \eqref{eq:EvolutionOfU2} in terms of the mixing parameters and 
their derivatives. 
Now we can solve for the derivatives of the mixing parameters.
Note that due to the separation of the evolution of the mass eigenvalues
in equation \eqref{eq:EvolutionOfEffectiveMasses}, we have reduced the
number of parameters from 12 to 9. 
The discussion so far has been very similar to the one of \cite{Casas:1999tg}.
There, the RG evolution of the mixing parameters is expressed
in terms of the mixing matrix elements and \(P'\). 

In order to obtain rather short and more explicit formulae, 
which are e.g.\ useful for deriving the approximations of 
Sec.~\ref{sec:AnalyticFormulae}, we now
consider \eqref{eq:EvolutionOfU2} and label the mixing parameters as
\begin{equation}
 \{\xi_k\} 
 = 
 \{\theta_{12},\theta_{13},\theta_{23},
        \delta,\delta_e,\delta_\mu,\delta_\tau,
        \varphi_1,\varphi_2\}
 \;.    
\end{equation}
We observe that the left-hand side of \eqref{eq:EvolutionOfU2}
is linear in \(\Dot\xi_k\).
Therefore, by solving the corresponding system of linear equations,
we can express the derivatives of the mixing parameters by
the mixing parameters, the mass eigenvalues and the Yukawa couplings.
The resulting formulae are still too long to be presented here 
but can be obtained from the web page 
\texttt{http://www.ph.tum.de/\textasciitilde{}mratz/AnalyticFormulae/}.

Finally, let us record that only the moduli of $U_{ij}$ enter into
the diagonal elements of
$P'$, if $P$ is diagonal, $P = \diag(P_1,P_2,P_3)$ (which is the case in
the SM and MSSM in the basis we have used in the main part), since
\begin{equation} \label{eq:PiiPrimeReal}
 P_{ii}' =
 \sum_{jk} (U^\dagger)_{ij} P_{jk} U_{ki} =
 \sum_{jk} U^*_{ji} P_j \delta_{jk} U_{ki} =
 \sum_j |U_{ji}|^2 P_j \;.
\end{equation}
Consequently, the evolution of the mass eigenvalues does not directly
depend on the Majorana phases, as claimed in
Sec.~\ref{sec:RunningOfMasses}.

\providecommand{\bysame}{\leavevmode\hbox to3em{\hrulefill}\thinspace}

\end{document}